\documentclass[11pt,11pt]{article}
\usepackage[pdftex]{graphicx}
\usepackage{amssymb}
\usepackage{amsmath}
\usepackage[]{enumerate}
\usepackage{bm}

\newcommand{\me}{\mathrm{e}}

\newcommand{\dif}{\mathrm{d}}
\newcommand{\al}{\alpha}
\newcommand{\bt}{\beta}

\newcommand{\p}{\partial}
\newcommand{\rv}{\bar{\bv}}

\DeclareBoldMathCommand{\bV}{V}
\DeclareBoldMathCommand{\bv}{v}
\DeclareBoldMathCommand{\bF}{F}
\DeclareBoldMathCommand{\bg}{g}
\DeclareBoldMathCommand{\bl}{\ell}
\DeclareBoldMathCommand{\bu}{u}
\DeclareBoldMathCommand{\br}{r}
\DeclareBoldMathCommand{\bx}{x}
\DeclareBoldMathCommand{\bg}{g}
\DeclareBoldMathCommand{\bb}{b}
\DeclareBoldMathCommand{\be}{e}
\DeclareBoldMathCommand{\bs}{s}
\DeclareBoldMathCommand{\bA}{A}
\DeclareBoldMathCommand{\bB}{B}
\DeclareBoldMathCommand{\bC}{C}
\DeclareBoldMathCommand{\bD}{D}
\DeclareBoldMathCommand{\bE}{E}
\DeclareBoldMathCommand{\bI}{I}
\DeclareBoldMathCommand{\bJ}{J}
\DeclareBoldMathCommand{\bM}{M}
\DeclareBoldMathCommand{\bL}{L}
\DeclareBoldMathCommand{\bN}{N}
\DeclareBoldMathCommand{\bP}{P}
\DeclareBoldMathCommand{\bR}{R}
\DeclareBoldMathCommand{\bS}{S}
\DeclareBoldMathCommand{\bU}{U}
\DeclareBoldMathCommand{\bW}{W}
\DeclareBoldMathCommand{\bk}{a}
\DeclareBoldMathCommand{\ba}{a}
\DeclareBoldMathCommand{\bn}{n}
\DeclareBoldMathCommand{\bp}{p}
\DeclareBoldMathCommand{\bq}{q}
\DeclareBoldMathCommand{\br}{r}
\DeclareMathOperator{\erf}{erf}

\begin{document}
\title{Coulomb collision operator}
\author{R. D. Hazeltine}
\date{\today}
\maketitle

\abstract Properties of the Coulomb collision operator that have been found useful in plasma kinetic theory and plasma transport theory are gathered and reviewed, with derivations.  The use of velocity-space coordinates appropriate to a magnetized (drift-kinetic) plasma is emphasized.  Little of the material is new, although the derivations may be more detailed than is customary.
\newpage
\tableofcontents
\newpage

\section{Operator} \label{sec:lb}
\subsection{Landau-Boltzmann form}
The distribution function for plasma species $a$ is denoted by $f_{a}(\bv)$. We use Boltzmann notation
\[
f'_{b} \equiv f_{b}(\bv')
\]
Then the Coulomb collision operator acting on species $a$ is
\[
C_{a} = \sum_{b}C_{ab}
\]
where\cite{braginskii}
\begin{equation}
\label{c0}
C_{ab} = \frac{\gamma_{ab}}{2} \frac{\partial}{\partial v_{\alpha}}\int d^{3}v\,' U_{\alpha \bt}\left(f'_{b}\frac{\partial f_{a}}{\partial v_{\bt}} - \frac{m_{a}}{m_{b}}f_{a}\frac{\partial f'_{b}}{\partial v'_{\bt}}\right)
\end{equation}
Here we have introduced the \emph{scattering tensor}
\begin{equation}
\label{defU}
U_{\alpha \bt} = \frac{u^{2}\delta_{\alpha \bt} - u_{\alpha}u_{\bt}}{u^{3}}, \,\,\,u_{\alpha} \equiv v_{\alpha} - v'_{\al}
\end{equation}
and the constant 
\[
\gamma_{ab} \equiv \frac{4\pi e_{a}^{2}e_{b}^{2}\ln \Lambda}{m_{a}^{2}}
\]
The charge and mass of species $a$ are $e_{a}$ and $m_{a}$; $\ln \Lambda$ is the Coulomb logarithm.  In this and the following Section the components $v_{\al}$ are Cartesian; curvilinear coordinates are introduced in Section \ref{curve}.

\subsection{Coefficient}
In terms of Braginskii's\cite{braginskii} electron-ion collision time $\tau_{ei}$, we have
\begin{equation}
\label{gt}
\gamma_{ei} = \frac{3\sqrt{\pi}v_{te}^{3}}{4n_{i}\tau_{ei}}
\end{equation}
where $n_{i}$ is the ion density and
\begin{equation}
\label{defvt}
v_{ta} \equiv \sqrt{2T_{a}/m_{a}}
\end{equation}
is the thermal speed.  Similarly,  in terms of the ion-ion collision time $\tau_{i}$,
\[
\gamma_{ii} = Z^{2}(m_{e}/m_{i})^{2}\gamma_{ei} = \frac{3\sqrt{2\pi}v_{ti}^{3}}{4n_{i}\tau_{i}}
\]
where $Z$ is the ionic charge.

The above formulae are in Gaussian units.  In SI units,
\[
\gamma_{ab} = \frac{e_{a}^{2}e_{b}^{2}\ln \Lambda}{4\pi \epsilon_{0}^{2}m_{a}^{2}}
\]
Some texts\cite{meiss92, framework} define $\gamma$ with different factors of $2$ and $m$. 

\subsection{Scattering and diffusion tensors}
The definition (\ref{defU}) implies that
\begin{eqnarray}
u_{\al}U_{\al \bt} &=& 0\label{uu0}\\
U_{\al\al} &=& 2/u \label{uaa}
\end{eqnarray}
The scattering tensor also satisfies
\begin{eqnarray}
U_{\al\bt}& = &\frac{\partial^{2}u}{\partial u_{\al}\partial u_{\bt}} = \frac{\partial}{\partial u_{\al}}\left(\frac{u_{\bt}}{u}\right) \label{id1}\\
\frac{\partial U_{\al\bt}}{\partial u_{\bt}} & = &  2 \frac{\partial}{\partial u_{\al}}\left(\frac{1}{u}\right) = \frac{\partial U_{\bt\bt}}{\partial u_{\al}}\label{id2}\\
\frac{\partial^{2}U_{\al\bt}}{\partial u_{\al}\partial u_{\bt}} & = & -8\pi \delta(\bu)\label{id3}
\end{eqnarray}

The averaged scattering tensor is the \emph{diffusion tensor}
\begin{equation}
\label{defd}
D_{b\, \al\bt} \equiv \int d^{3}v\,' f'_{b} U_{\al\bt}
\end{equation}
We introduce the Maxwellian distribution
\begin{equation}
\label{deffm}
f_{Mb} \equiv \frac{n_{b}}{\pi^{3/2}v_{tb}^{3}} e^{-v^{2}/v_{tb}^{2}}
\end{equation}
and consider the Maxwellian diffusion tensor
\[
D_{Mb\,\al\bt} \equiv \int d^{3}v\,' f'_{Mb} U_{\al\bt}
\]
In terms of the variable $s_{\al} \equiv v_{\al}/v_{tb}$ one finds that
\begin{equation}
\label{dm}
D_{Mb\,\al\bt} = \frac{n_{b}}{2v_{tb}}\left[\delta_{\al\bt}M_{1}(s) - s_{\al}s_{\bt}M_{2}(s)\right]
\end{equation}
with
\begin{eqnarray}
M_{1} &=& s^{-3}[s\, \erf'(s) + (2s^{2} - 1)\erf(s)]\label{defm1} \\
M_{2} &=& 3s^{-5}[s\, \erf'(s) + (2s^{2}/3 - 1)\erf(s)] = -s^{-1}dM_{1}/ds \label{defm2}
\end{eqnarray}
Here
\[
\erf(s) = \frac{2}{\sqrt{\pi}}\int_{0}^{s}dt e^{-t^{2}}
\]
denotes the error function and $\erf'(s) \equiv d\erf /ds$.  The relation
\begin{equation}
\label{m1m2}
s^{3}(M_{1} - s^{2}M_{2}) = 2 (\erf - s\erf')
\end{equation}
is often useful. The gradient of the Maxwellian diffusion tensor is surprisingly simple:
\begin{equation}
\label{dmd}
\frac{\p D_{Mb\, \al\bt} }{\p v_{\bt}} = 2\frac{n_{b}}{v_{tb}^{2}}\frac{s_{\al}}{s}\frac{\p}{\p s}\left(\frac{\erf(s)}{s}\right)
\end{equation}
A derivation of this relation appears in Subsection \ref{subsec:rmj}.

From the well-known formulae\cite{abram},
\begin{eqnarray*}
\erf(s) &=& 1 +\mathcal{O}(e^{-s^{2}}), \;\;\mbox{for} \,\,s \gg 1\\
 &=& \frac{2}{\sqrt{\pi}}\left(s - \frac{s^{3}}{3} + \frac{s^{5}}{10} - \cdots \right), \;\;\mbox{for}\,\,s\ll 1
\end{eqnarray*}
we obtain the asymptotic forms
\begin{eqnarray}
M_{1} &\approx& 2s^{-1}, \;\;\; M_{2} \approx 2s^{-3}, \,\,\mbox{for}\,\, s \gg 1,\label{asym1}\\
M_{1} &\approx& 8/(3\sqrt{\pi}), \;\;\; M_{2} \approx 16/(15\sqrt{\pi})\,\,\mbox{for}\,\, s \ll 1,\label{asym2}
\end{eqnarray}

\subsection{Fokker-Planck (FP) form}
We integrate by parts in (\ref{c0}) and use 
$\partial u/\partial \bv'  = -\partial u/\partial \bv$ to obtain
\begin{equation}
\label{cfp}
C_{ab} =  \frac{\gamma_{ab}}{2}\frac{\partial}{\partial v_{\al}}\left(D_{b\, \al\bt}\frac{\partial f_{a}}{\partial v_{\bt}} - \frac{m_{a}}{m_{b}}f_{a} \frac{\partial }{\partial v_{\bt}} D_{b\,\al\bt}\right) 
\end{equation}
This is standard Fokker-Planck form; the first parenthesized term represents diffusion and the second represents dynamical friction.  An equivalent expression is
\begin{equation}
\label{cfp2}
C_{ab} =  \frac{\gamma_{ab}}{2}\frac{\partial}{\partial v_{\al}}\left[\frac{\partial}{\partial v_{\bt}}\left(f_{a}D_{b\, \al\bt}\right) - \left(1 + \frac{m_{a}}{m_{b}}\right) f_{a} \frac{\partial D_{b\,\al\bt}}{\partial v_{\bt}} \right]
\end{equation}

\subsection{Rosenbluth-MacDonald-Judd (RMJ) form}\label{subsec:rmj}

We introduce the functions
\begin{eqnarray}
G_{b} &\equiv & \int d^{3}v\,' f'_{b}\, u \label{defg}\\
H_{b} &\equiv & \int d^{3}v\,' f'_{b}\, u^{-1}\label{defh}
\end{eqnarray}
and observe from (\ref{id1}) and (\ref{id2}) that
\begin{eqnarray}
D_{b\, \al\bt} & = & \frac{\partial^{2}G_{b}}{\partial v_{\al}\partial v_{\bt}}\label{dg} \\
\frac{\partial D_{b\, \al\bt}}{\partial v_{\bt}}  & = & 2\frac{\partial H_{b}}{\partial v_{\al}} \label{dh}
\end{eqnarray}
Therefore (\ref{cfp2}) can be written as
\begin{equation}
\label{rmj}
C_{ab} =  \frac{\gamma_{ab}}{2}\frac{\partial}{\partial v_{\al}}\left[\frac{\partial}{\partial v_{\bt}}\left(f_{a} \frac{\partial^{2}G_{b}}{\partial v_{\al}\partial v_{\bt}} \right) -2\left(1 + \frac{m_{a}}{m_{b}}\right) f_{a}\frac{\partial H_{b}}{\partial v_{\al}} \right]
\end{equation}
The original RMJ paper\cite{RMJ} expressed the \emph{full} collision operator $C_{a}$ for a two-species plasma, in terms of species-sums of $G_{b}$ and $H_{b}$.  We find the individual species version more convenient, and identify $G_{b}$ and $H_{b}$ with the so-called Rosenbluth potentials.  The name stems from the identities
\begin{eqnarray}
\nabla_{\bv}^{2}G_{b} \equiv \frac{\partial^{2}G_{b}}{\partial v_{\al}\partial v_{\al}} &=& 2 H_{b}\label{lgh}\\
\nabla_{\bv}^{2}H_{b} = -4\pi f_{b}\label{d2h}
\end{eqnarray}
which follow from (\ref{uaa}) and (\ref{id3}), and which resemble Poisson's equation for the electrostatic potential.

Noticing that $2H_{b}$ is a source for $G_{b}$ in the same way that $-4\pi f_{b}$ is a source for $H_{b}$, we conclude from (\ref{defh}) that
\begin{equation}
\label{gib}
G_{b} = -\frac{1}{2\pi}\int \dif^{3}v' H'_{b}u^{-1}
\end{equation}
This relation is sometimes more useful than (\ref{defg}).

We use an $M$-subscript to distinguish the potentials evaluated on a Maxwellian distribution: $G_{Mb} = G_{b}( f_{Mb})$, etc.  One finds 
\begin{equation}
G_{Mb}  = (n_{b}v_{tb}/2)[ \erf'(s) + s^{-1}(1 + 2s^{2})\erf(s)]\label{gm}
\end{equation}
where $s = v/v_{tb}$ as usual.  Similarly
\begin{equation}
\label{hm}
H_{Mb} = (n_{b}/v_{tb})s^{-1}\erf(s)
\end{equation}
Note that substitution of (\ref{hm}) into (\ref{dh}) yields (\ref{dmd}).

\section{Symmetry of the collision operator}
\subsection{Scalar property}
We consider here transformations in velocity space,
\[
\bv \rightarrow \rv
\]
We are especially interested in \emph{passive} transformations, for which $\bv$ and $rv$ refer to the same point; only the coordinate system is transformed.  However any passive transformation also has an \emph{active} interpretation, for which $\bv$ and $\rv$ refer to different points.  

A coordinate change will in general affect the form of functions and operators, related by a \emph{transformation law}, 
\[
f \rightarrow \bar{f}, C \rightarrow \bar{C}
\]
Two types of transformation have special special importance:  rotations and Galilean transformations.  We will refer to a Galilean trans\-for\-ma\-tion as a ``boost''.  In re\-la\-tivistic theory, ro\-ta\-tions and boosts are simply two types of homo\-geneous Lorentz transformations, and both can be treated simultaneously.  However in non-relativistic theory they are distinct; in particular, rotations are linear and homogeneous, while boosts are inhomogeneous (affine).  

The simplest transformation law is that of a \emph{scalar} function $s(\bv)$:
\begin{equation}
\label{defscalar}
\bar{s}(\rv) = s(\bv)
\end{equation}
We expect any physically meaningful quantity without coordinate indices to be scalar, under both rotations and boosts. In particular it can be shown that a distribution function is a scalar.  Recall also that the full contraction of any even-rank tensor is necessarily a scalar under rotation; for example, if $T_{\al\bt}$ is  a tensor under rotation, then the quantity
\[
s = v_{\al}T_{\al\bt}v_{\bt}
\]
is a scalar.

The collision operator is a \emph{scalar operator}.  To make this nomenclature explicit, we write the operator as
\[
C_{ab}= C_{ab}[f_{a}, f_{b}](\bv)
\]
It is thus a function\emph{al} of the two distributions, but an ordinary function of velocity $\bv$.  The transformed operator is
\[
 \bar{C}_{ab}[\bar{f}_{a}, \bar{f}_{b}](\rv)
 \]
The scalar nature of $C_{ab}$ is therefore expressed by:
\begin{equation}
\label{punch}
\bar{C}_{ab}[\bar{f}_{a}, \bar{f}_{b}](\rv) = C_{ab}[f_{a}, f_{b}](\bv)
\end{equation}
This relation is to hold for both rotations and boosts in velocity space.

\subsection{Invariance}\label{subsec:inv}
A quantity $q(\bv)$ is \emph{invariant} under some transformation $\bv \rightarrow \rv$ if it satisfies
\begin{equation}
\label{definv}
q(\bv) = q(\rv)
\end{equation}
For example, any function that depends on velocity only through is magnitude $v = |\bv|$ is invariant under rotation.  Invariance is often best understood in terms of active transformations: an invariant quantity is the same at the points $\bv$ and $\rv$. 


Being a scalar and being invariant are entirely independent properties. In particular (\ref{definv}) makes no reference to $\bar{q}$, so its transformation law is irrelevant.  

A quantity that is invariant under rotation is called rotationally symmetric. A function that is rotationally symmetric has the same value at any two points that can be connected by a rotation.
Distribution functions that are rotationally symmetric are called \emph{isotropic}; although the Maxwellian distribution is isotropic, most distribution functions are not. 

The collision operator is both rotationally symmetric and invariant under Galilean transformations.  That is, for either type of transformation, 
\begin{equation}
\label{cinv}
C_{ab}[f_{a}, f_{b}](\bv) = C_{ab}[\bar{f}_{a}, \bar{f}_{b}](\rv)  
\end{equation}
Verifying (\ref{cinv}) is the main point of this Section.

Note that symmetry of the operator, in the sense of (\ref{cinv}), does not imply symmetry of the velocity-function $C_{ab}(\bv)$, because the distributions are not in general symmetric.  

\subsection{Rotations}
We denote the rotation matrix by $\bR$, so  that the basic transformation law is
\[
\rv = \bR \cdot \bv
\]

Recall that the matrix inverse to $\bR$ is the transposed matrix $\bR^{T}$:
\begin{equation}
\label{rr1}
\bR^{T}\cdot \bR =\bm{I}
\end{equation}
where $\bm{I}$ is the identity matrix.  Thus
\begin{equation}
\label{trinv}
\bv = \bR^{T}\cdot \rv
\end{equation}
and 
\[
\frac{\p}{\p v_{\al}} = \frac{\p \bar{v}_{\lambda}}{\p v_{\al}} \frac{\p}{\p \bar{v}_{\lambda}} = R_{\lambda \al} \frac{\p}{\p \bar{v}_{\lambda}}
\]
That is
\begin{equation}
\label{tpart}
\frac{\p}{\p \bv} = \bR^{T}\cdot \frac{\p}{\p \rv}
\end{equation}
That $\bv$ and $\p /\p \bv$ transform identically follows from the fact that the metric is the identity matrix: we are using Cartesian coordinates.

The scalar transformation law noted in (\ref{defscalar}) can now be written more explicitly as
\begin{equation}
\label{trf}
f(\bR^{T}\cdot \rv) = \bar{f}(\rv)
\end{equation}
Similarly the scattering tensor $\bU$ transforms according to
\[
\bar{\bU}(\bar{\bu}) = \bR\cdot \bU(\bu)\cdot \bR^{T}
\]
or
\begin{equation}
\bU(\bR^{T}\cdot \bar{\bu}) = \bR^{T}\cdot \bar{\bU}(\bar{\bu})\cdot \bR \label{tru}
\end{equation}

Any quantity constructed by full contraction of higher-rank tensors will automatically transform under rotation as a scalar.  Thus (\ref{c0}) shows immediately that the collision operator is a scalar under rotation:  the scalar nature of $f$ makes $\p f/\p \bv$ a vector,  the scattering tensor is indeed a tensor, and so on.  

Another property of $C_{ab}$ is also clear from  (\ref{c0}): the operator is rotationally symmetric, in the sense of (\ref{cinv}).  The point is that the operator contains no information that distinguishes a particular direction in velocity space.  Therefore observers of collisional effects with different orientations would see identical processes.  

Despite the transparency of rotational symmetry, we next verify this property in detail, by expressing the right-hand side of (\ref{c0}) in terms of barred quantities. Notice that no transformation property of $C_{ab}$  is involved, only its expression in terms of different coordinates.  To express the collision operator in terms of $\rv$, we proceed as follows: \begin{enumerate} \item Express $\bv$ in terms of $\rv$, introducing the factors of $\bR^{T}$ as required by (\ref{trinv}) and (\ref{tpart}). \item Similarly transform the dummy variable $\bv'$, noting that $d^{3}v' = d^{3}\bar{v}'$.  Thus all velocities $\bv$ and $\bv'$ are eliminated from $C_{ab}$, in favor of $\rv$ and $\rv'$. \item Use (\ref{trf}) and (\ref{tru}) to express the distribution functions and scattering tensor in terms of the corresponding barred quantities.  \end{enumerate}

After this step the collision operator has the form
\[\begin{split}
C_{ab} =& \frac{\gamma_{ab}}{2}\bR^{T}\cdot \frac{\p}{\p \rv}\cdot\int d^{3}\bar{v}' \, \bR^{T}\cdot \bar{\bU}\cdot \bR \\
&\times \left[\bar{f}_{b}(\rv')\bR^{T}\cdot \frac{\p \bar{f}_{a}(\rv)}{\p \rv} - \frac{m_{a}}{m_{b}}\bar{f}_{a}(\rv)\bR^{T}\cdot \frac{\p \bar{f}_{b}(\rv')}{\p \rv'}\right]\end{split}
\]
or, in view of (\ref{rr1}), 
\begin{equation}\label{crot1}\begin{split}
C_{ab} =& \frac{\gamma_{ab}}{2}\bR^{T}\cdot \frac{\p}{\p \rv}\cdot\int d^{3}\bar{v}' \, \bR^{T}\cdot \bar{\bU}\\
&\times \left[\bar{f}_{b}(\rv') \frac{\p \bar{f}_{a}(\rv)}{\p \rv} - \frac{m_{a}}{m_{b}}\bar{f}_{a}(\rv)  \frac{\p \bar{f}_{b}(\rv')}{\p \rv'}\right]\end{split}
\end{equation}
To confirm that the remaining factors of $\bR^{T}$ cancel, we consider the vector 
\begin{eqnarray*}
\bR^{T}\cdot \frac{\p}{\p \rv}\cdot\int d^{3}\bar{v}' \, \bR^{T}\cdot \bar{\bU} 
&=&R_{\lambda \kappa}\frac{\p}{\p \bar{v}_{\lambda}}\int d^{3}\bar{v}'R_{\al\kappa}\bar{U}_{\al\bt}  \\
&=&R_{\lambda \kappa}R_{\al\kappa}\frac{\p}{\p \bar{v}_{\lambda}}\int d^{3}\bar{v}'\, \bar{U}_{\al\bt} \\
&=&\delta_{\lambda \al}\frac{\p}{\p \bar{v}_{\lambda}}\int d^{3}\,\bar{v}' \bar{U}_{\al\bt} \\
&=& \frac{\p}{\p \rv}\cdot\int d^{3}\bar{v}'\, \bar{\bU} 
\end{eqnarray*}
Therefore
\begin{equation}\label{crot2} 
C_{ab} = \frac{\gamma_{ab}}{2}  \frac{\p}{\p \rv} \cdot\int d^{3}\bar{v}'\, \bar{\bU}
\cdot \left[\bar{f}_{b}(\rv') \frac{\p \bar{f}_{a}(\rv)}{\p \rv} - \frac{m_{a}}{m_{b}}\bar{f}_{a}(\rv)\cdot \frac{\p \bar{f}_{b}(\rv')}{\p \rv'}\right] 
\end{equation}
Observing that the right-hand side of (\ref{crot2}) coincides with $C_{ab}[\bar{f}_{a},\bar{f}_{b}](\rv)$, we conclude that (\ref{cinv}) holds:  the collision operator is invariant under rotation.  

It is clear that the nature of the scattering tensor $\bU$, depending on $\bv$ and $\bv'$ alone, is crucial to this result. For example the operator obtained from $C_{ab}$ by the replacement
\[
\bU \rightarrow \bk\bk
\]
where $\bk$ is some fixed vector, such as the direction of an external field, would be a scalar like $C_{ab}$, but not invariant.

\subsection{Boosts}
Here we consider the coordinate transformation
\[
\rv = \bv - \bV
\]
where $\bV$ is some fixed velocity.  Evidently the components $\bv$ and $\rv$ are those seen in two frames with relative velocity $\bV$; again we emphasize that both sets of coordinates refer to the same point in velocity space.  As before we can define $\bar{C}_{ab}$ to insure that the operator is a scalar under Galilean boosts.  The more interesting point is its invariance under such boosts:  the collisional process is described by the same operator in any inertial frame.  This fact is not surprising but nonetheless we verify it in detail.

The scalar nature of the distribution is expressed by $\bar{f}(\rv) = f(\bv)$, or
\begin{equation}
\label{defgalf}
\bar{f}_{a}(\rv) = f_{a}(\rv + \bV) 
\end{equation}
Derivatives of the distributions satisfy $\p/\p \rv = \p/\p \bv$ and the scattering tensor satisfies
\[
\bU(\rv, \rv') \equiv \bar{\bU} =  \bU(\bv, \bv')
\]
With these facts, we can proceed as in the rotational case, expressing the operator $C_{ab}$ in terms of the transformed velocity $\rv$ (without transforming the operator).  After changing the integration variable as usual, $\bv' \rightarrow \rv'$, one finds
\begin{equation}\begin{split}
&C_{ab}[f_{a}, f_{b}](\bv) = \frac{\gamma_{ab}}{2} \frac{\p}{\p \rv} \cdot\int d^{3}\bar{v}'\, \bar{\bU} \\ &
\times\left[f_{b}(\rv' + \bV) \frac{\p f_{a}(\rv+\bV)}{\p \rv} - \frac{m_{a}}{m_{b}}f_{a}(\rv+\bV)  \frac{\p f_{b}(\rv'+\bV)}{\p \rv'}\right]\end{split}\label{cgal}
\end{equation}
If we now use (\ref{defgalf}) to replace each distribution $f$ by its transformed version $\bar{f}$, we see that the right-hand side of (\ref{cgal}) becomes $C_{ab}[\bar{f}_{a}, \bar{f}_{b}](\rv)$. Hence (\ref{cinv}) pertains: the collision operator is invariant under Galilean transformations.  The key underlying fact here is the dependence of the scattering tensor $\bU$ on the velocity difference $\bv - \bv'$ alone.

\section{Covariant expression}\label{curve}
\subsection{Notation}
Next we express our key results in covariant form, valid for any velocity coordinates, $(\xi^{1}, \xi^{2}, \xi^{3})$.  The metric tensor is denoted by $g_{\al\bt}$:
\[
d\bv\cdot d\bv = g_{\al\bt}d\xi^{\al}d\xi^{\bt}
\]
and its determinant is denoted by $g$, so that the Jacobian is $\sqrt{g}$:
\[
d^{3}v\, = \sqrt{g}\, d^{3}\xi
\]
The inverse matrix is $g^{\al\bt}$.

Covariant derivatives are denoted by commas.  In particular, if $S$ is a velocity-space scalar, then
\begin{equation}
\label{sab}
S_{,\al,\bt} = \left(\frac{\partial S}{\partial \xi^{\al}}\right)_{,\bt} = \frac{\partial^{2}S}{\partial \xi^{\al}\partial \xi^{\bt}} - \Gamma^{\kappa}_{\al\bt} \frac{\partial S}{\partial \xi^{\kappa}}
\end{equation}
Here $\Gamma$ is the Christoffel symbol, given by
\[
\Gamma^{\kappa}_{\al\bt} = \frac{1}{2}g^{\kappa\lambda}\left(\frac{\partial g_{\lambda \bt}}{\partial \xi^{\al}} + \frac{\partial g_{\lambda \al}}{\partial \xi^{\bt}} - \frac{\partial g_{\al \bt}}{\partial \xi^{\lambda}}\right)
\]
A scalar that is often useful is the kinetic energy,
\[
w \equiv \frac{1}{2}\bv \cdot \bv
\]
Notice that
\begin{equation}
\label{vw}
v_{\al} = w_{,\al}
\end{equation}
is a covariant vector, for any coordinate system. It is important to appreciate that the contravariant quantities
\begin{equation}
\label{defvcontra}
v^{\al} = g^{\al\bt}w_{,\bt}
\end{equation}
are in general not coordinates in velocity space (the latter are denoted by $\xi^{\al}$).  The $v^{\al}$ are coordinates only in the Cartesian case.

The normalized version is of course $s_{\al} = w_{,\al}/v_{t}$ with contravariant components
\begin{equation}
\label{defscontra}
s^{\al}\equiv v_{t}^{-1}g^{\al\bt}w_{,\bt}
\end{equation}
 
A special case of (\ref{sab}) is 
\begin{eqnarray}
\nabla^{2}S &=&g^{\al\bt}S_{\al,\bt} \label{laplace}\\
&=& \frac{1}{\sqrt{g}}\frac{\p}{\p \xi^{\al}}\sqrt{g}\, g^{\al\bt}\frac{\p S}{\p \xi^{\bt}}\nonumber
\end{eqnarray}

\subsection{Covariant scattering and diffusion}
We begin with  the velocity-space scalar $u \equiv |\bv - \bv'|$, the distance between $\bv$ and a fixed, chosen point $\bv'$.  Then we can define the covariant vector $u_{\al}$ by
\[
u_{\al} = uu_{,\al}
\]
where the derivative is taken with respect to $\bv$ as usual.  Similarly, the covariant definition of $U_{\al\bt}$ is
\begin{equation}
\label{defbigu}
U_{\al\bt} \equiv u_{,\al,\bt}
\end{equation}
Notice that the indices of $U_{\al\bt}$ are raised by the metric elements evaluated at $\bv$, never at $\bv'$.  

Some useful identities follow from (\ref{uu0}) -- (\ref{id3}):
\begin{eqnarray}
u_{\kappa}g^{\kappa \al}U_{\al\bt} &=&  0 \\
U^{\al\bt}_{,\bt} &=& g^{\al\bt}(2/u)_{,\bt}\\
U^{\al\bt}_{,\al,\bt} &=& -8\pi\delta(\bu)/\sqrt{g}
\end{eqnarray}

These imply, as in (\ref{dg}) -- (\ref{dh}),
\begin{eqnarray}
D_{b\, \al\bt} &=& G_{b\, ,\al,\bt}\\\label{dg2}
D_{b,\al,\bt}^{\al\bt} &=& -8\pi f_{b}\label{ddf} \\
D_{b\, \al\, ,\bt}^{\bt} &=& 2 H_{b\, ,\al}  \label{hd}
\end{eqnarray}
Note also that the covariant version of (\ref{uaa}) implies
\begin{equation}
H_{b} = \frac{1}{2}g_{\al\bt} D_{b}^{\al\bt}\label{hgd}
\end{equation}

The vector given by (\ref{hd}) can be identified with dynamical friction; in fact a convenient notation is
\begin{equation}
\label{defdb}
\mathcal{F}_{ab}^{\al} =\left(1 + \frac{m_{a}}{m_{b}}\right)g^{\al\bt}H_{b\, ,\bt} = \frac{1}{2}\left(1 + \frac{m_{a}}{m_{b}}\right)D_{b\,,\bt}^{\al\bt}
\end{equation}

As usual we use an $M$-subscript to indicate evalua\-tion on a Max\-wellian dis\-tri\-bution.  The relation (\ref{dm}) has the covariant form
\begin{equation}
\label{dm2}
D_{Mb\, \al\bt} = \frac{n_{b}}{2v_{tb}}\left(g_{\al\bt}M_{1} - v_{tb}^{-2}w_{,\al}w_{,\bt}M_{2}\right)
\end{equation}
where the $M$'s are given by (\ref{defm1}) and (\ref{defm2}) as before.
This is indeed a covariant second-rank tensor, because $w$, $M_{1}$ and $M_{2}$ are all velocity-space scalars.

The Maxwellian diffusion tensor satisfies
\begin{equation}
\label{divd}
D_{Mb\, ,\al}^{\al\bt} = -\frac{m_{b}}{T_{b}}w_{,\al}D_{Mb}^{\al\bt}
\end{equation}
Thus the dynamical friction force in the Maxwellian case is
\begin{equation}
\label{fric2}
\mathcal{F}^{\al}_{Mab} = -\frac{m_{b}}{2T_{b}}\left(1 + \frac{m_{a}}{m_{b}}\right)w_{,\al}D_{Mb}^{\al\bt}
\end{equation}

\subsection{Covariant collision operator}
The covariant FP operator---the  covariant form of (\ref{cfp2})---is 
\begin{equation}
\label{cfp3}
C_{ab} =   \frac{\gamma_{ab}}{2} \left[\left(f_{a}D_{b}^{\al\bt}\right)_{,\bt} - 2f_{a}F_{ab}^{\al} \right]_{,\al}
\end{equation}
The more explicit version,
\begin{equation}
\label{cfp4}
C_{ab} =   \frac{\gamma_{ab}}{2}\left[8\pi \frac{m_{a}}{m_{b}}f_{a}f_{b} + f_{a\, ,\al,\bt}D_{b}^{\al\bt} + \left(1 - \frac{m_{a}}{m_{b}}\right) f_{a\, ,\bt}D_{b\, ,\al}^{\al\bt}\right]
\end{equation}
is sometimes useful.  Here we used (\ref{ddf}).

In terms of the Rosenbluth potentials we have
\begin{equation}
\label{rmj2}
C_{ab} = \frac{\gamma_{ab}}{2}\left[ \left(f_{a}g^{\al\kappa}g^{\bt\lambda}G_{b\, ,\kappa,\lambda}\right)_{,\bt} - 2\left(1 + \frac{m_{a}}{m_{b}}\right) f_{a}g^{\al\kappa}H_{b\,,\kappa}\right]_{,\al}
\end{equation}
or
\begin{equation}
\label{rmj3}\begin{split}
C_{ab} =&   \frac{\gamma_{ab}}{2}\left[8\pi \frac{m_{a}}{m_{b}}f_{a}f_{b} +  f_{a\, ,\al,\bt}g^{\al\kappa}g^{\bt\lambda}G_{b\, ,\kappa,\lambda}\right. \\ &\left.+2\left(1 - \frac{m_{a}}{m_{b}}\right)f_{a\, ,\kappa}g^{\kappa\lambda}H_{b\, ,\lambda}\right]\end{split}
\end{equation}

\section{Coordinates for magnetized plasma}
\subsection{Spherical polar coordinates}\label{subsec:spher}
These are the coordinates used by RMJ. We begin with
\begin{equation}
\label{v1}
\bv = \bb v_{\parallel} + v_{\perp}(\be_{1}\cos \gamma +\be_{2}\sin \gamma )
\end{equation}
Here the unit vectors $(\be_{1}, \be_{2}, \bb)$ form a local  positive triplet, with $\bb$ denoting the direction of the magnetic field; $\gamma$ is the gyrophase angle; and the remaining notation is self-explanatory. We express the parallel and perpendicular speeds in terms of the Cosine of the pitch angle, $\xi \equiv v_{\parallel}/v$:
\[
v_{\parallel} = v\xi, \,\,\, v_{\perp}  = v\sqrt{1 - \xi^{2}}
\]
By spherical velocity coordinates, we mean the set
\[
(\xi_{1}, \xi_{2}, \xi_{3}) = (v, \xi, \gamma)
\]
The corresponding metric tensor is diagonal (i.e., the coordinates are orthogonal), with 
\[
g_{11} = 1,\;\; g_{22} = v^{2}/(1 - \xi^{2}), \;\; g_{33} = v^{2}(1 - \xi^{2})
\]
and therefore $\sqrt{g} = v^{2}$.  Because $g_{\al\bt}$ is diagonal, its inverse $g^{\al\bt}$ is also diagonal, with
\[
g^{\al\al} = \frac{1}{g_{\al\al}} \;\;\;\mbox{(no sum)}
\]

The only non-vanishing Christoffel symbols are:
\begin{eqnarray*}
&\Gamma^{1}_{22}& = -v/(1 - \xi^{2}), \;\; \Gamma^{1}_{33} = -v(1 - \xi^{2})\\
&\Gamma^{2}_{22}& = \xi/(1 - \xi^{2}), \;\; \Gamma^{2}_{33} = \xi (1 - \xi^{2})\\
&\Gamma^{2}_{12}& =\Gamma^{2}_{21}= 1/v, \;\; \Gamma^{3}_{23}= \Gamma^{3}_{32} = -\xi/(1 - \xi^{2}), \;\;\Gamma^{3}_{13}=\Gamma^{3}_{31} = 1/v
\end{eqnarray*}
We apply these formulae to the tensor  $S_{,\al,\bt}$ of (\ref{sab}) and find
\begin{eqnarray*}
&S_{,1,1}& = \frac{\p^{2}S}{\p v^{2}}, \\
&S_{,2,2}& = \frac{\p^{2}S}{\p \xi^{2}} + \frac{v}{1 - \xi^{2}}\frac{\p S}{\p v} - \frac{\xi}{1 - \xi^{2}}\frac{\p S}{\p \xi},\\
&S_{,3,3}& = \frac{\p^{2} S}{\p \gamma^{2}} + v(1 - \xi^{2})\frac{\p S}{\p v} - \xi (1 - \xi^{2}) \frac{\p S}{\p \xi}\\
&S_{,1,2}& = S_{,2,1} = \frac{\p^{2}S}{\p v\p \xi} - \frac{1}{v}\frac{\p S}{\p \xi}\\
&S_{,1,3}& = S_{,3,1} = \frac{\p^{2}S}{\p v\p \gamma}- \frac{1}{v}\frac{\p S}{\p \gamma}\\
&S_{,2,3}& = S_{,3,2} = \frac{\p^{2}S}{\p \xi \p \gamma} + \frac{\xi}{1 - \xi^{2}}\frac{\p S}{\p \gamma}
\end{eqnarray*}
We can combine these results with (\ref{laplace}) to compute
\begin{equation}
\nabla^{2}S 
= \frac{1}{v^{2}}\frac{\p}{\p v}v^{2}\frac{\p S}{\p v} + \frac{1}{v^{2}}\frac{\p}{\p \xi}(1 - \xi^{2})\frac{\p S}{\p \xi} + \frac{1}{v^{2}(1 - \xi^{2})}\frac{\p^{2} S}{\p \gamma^{2}}\label{del2spher}
\end{equation}

A simpler quantity is
\[
v^{\al}S_{,\al} = g^{\al \bt}\frac{\p (v^{2}/2)}{\p \xi^{\bt}}\frac{\p S}{\p \xi^{\al}}
\]
Here we used (\ref{vw}) and (\ref{defvcontra}).  Since
\[
\frac{\p (v^{2}/2)}{\p \xi^{\bt}} = v\delta^{1}_{\bt}
\]
we have
\begin{equation}
\label{vds}
v^{\al}S_{,\al} = v\frac{\p S}{\p v}
\end{equation}

\subsection{Quasi-invariant coordinates}\label{subsec:qic}
When the electrostatic potential changes slowly along the guiding center orbit, the kinetic energy $w = v^{2}/2$ is nearly constant.  Similarly for small gyroradius the magnetic moment $v_{\perp}^{2}/2B$ is approximately constant, where $B$ is the magnetic field magnitude.  Therefore the coordinates
\begin{equation}
\label{qic}
(\xi^{1}, \xi^{2}, \xi^{3}) = (w, \lambda, \gamma)
\end{equation}
are often convenient.  Here $\lambda \equiv v_{\perp}^{2}/(2wB)$ is called the pitch-angle variable.  Then
\[
v_{\parallel} = \sigma \sqrt{2 w (1 - \lambda B)}, \;\; v_{\perp} = \sqrt{2w\lambda B}
\]
and the non-vanishing components of the metric tensor are
\[
g_{11} = \frac{1}{2w}, \;\; g_{22} = \frac{wB}{2\lambda (1 - \lambda B)}, \;\; g_{33} = 2w\lambda B
\]
whence the Jacobian
\[
\sqrt{g} = B\sqrt{\frac{w}{2(1 - \lambda B)}} = \frac{Bw}{|v_{\parallel}|}
\]
Again the inverse tensor is trivial:
\[
g^{\al\al} = \frac{1}{g_{\al\al}} \;\;\;\mbox{(no sum)}
\]
The non-vanishing Christoffel symbols are
\begin{eqnarray*}
&\Gamma^{1}_{11}& = -\frac{1}{2w}, \;\;\Gamma^{1}_{22} = -\frac{wB}{2\lambda(1 - \lambda B)},\;\; \Gamma^{1}_{33} = -2w\lambda B \\
&\Gamma^{2}_{22}& = -\frac{1 - \lambda B}{2\lambda (1 - \lambda B)},\;\; \Gamma^{2}_{33} = -2\lambda (1 - \lambda B),\\
&\Gamma^{3}_{13}& = \Gamma^{3}_{31}=\frac{1}{2w},\;\; \Gamma^{3}_{23} =  \Gamma^{3}_{32}=\frac{1}{2\lambda}, \;\;\Gamma^{2}_{12} = \Gamma^{2}_{21} = \frac{1}{2w},
\end{eqnarray*}
and the tensor $S_{,\al ,\bt}$ has the following components:
\begin{eqnarray*}
&S_{,1,1}& = \frac{\p^{2}S}{\p w^{2}} + \frac{1}{2w}\frac{\p S}{\p w} \\
&S_{,2,2}& = \frac{\p^{2}S}{\p \lambda^{2}}+ \frac{1}{2\lambda(1 - \lambda B)}\left[(1 - 2\lambda B)\frac{\p S}{\p \lambda} + wB\frac{\p S}{\p w}\right]\\
&S_{,3,3}& = \frac{\p^{2}S}{\p \gamma^{2}} + 2w\lambda B \frac{\p S}{\p w} + 2\lambda (1 - \lambda B) \frac{\p S}{\p \lambda}\\
&S_{,1,2}& = S_{,2,1} = \frac{\p^{2}S}{\p \lambda \p w} - \frac{1}{2w}\frac{\p S}{\p \lambda},\\
&S_{,1,3}& = S_{,3,1} = \frac{\p^{2}S}{\p w \p \gamma} - \frac{1}{2w}\frac{\p S}{\p \gamma}\\
&S_{,2,3}& = S_{,3,2} = \frac{\p^{2}S}{\p \lambda \p \gamma} - \frac{1}{2\lambda}\frac{\p S}{\p \gamma}
\end{eqnarray*}

We can substitute these expressions into (\ref{laplace}) to compute
\begin{equation}\begin{split}
\nabla^{2}S =&\frac{2\sqrt{1 - \lambda B}}{wB}\frac{\p}{\p \lambda}\lambda\sqrt{1 - \lambda B}\frac{\p S}{\p \lambda}\label{del2s}\\&+ \frac{2}{w^{1/2}}\frac{\p}{\p w}w^{3/2}\frac{\p S}{\p w} + \frac{1}{2w\lambda B}\frac{\p^{2}S}{\p \gamma^{2}}\end{split}
\end{equation}

These formulae, or those of Subsection \ref{subsec:spher}, can be directly substituted into (\ref{rmj3}), with $S = f_{a}$ and $S = G_{b}$.  Thus we obtain an explicit expression for the general nonlinear collision operator in the two magnetized-plasma coordinate systems.

As a simple application of quasi-invariant coordinates, we verify (\ref{divd}). According to (\ref{hd}) it's left-hand side is (suppressing the species-subscript)
\[
D_{M\, ,\bt}^{\al\bt} = g^{\al\kappa}D^{\bt}_{M\, \kappa,\bt} = 2g^{\al\kappa}H_{M\, ,\kappa}
\]
In quasi-invariant coordinates, the covariant deriviative of $H_{M}$ has only a single component, $H_{M\, ,\kappa} = \delta_{\kappa\, 1}dH_{M}/dw$, and 
\[
\frac{d}{dw} = \frac{1}{v_{t}^{2}s}\frac{d}{ds}
\]
so (\ref{hm}) gives
\begin{eqnarray}
D_{M\, ,\bt}^{\al\bt} &=& \frac{2n}{v_{t}}g^{\al\kappa}\delta_{\kappa\, 1} \frac{1}{v_{t}^{2}s}\frac{d}{ds}\left(\frac{\erf}{s}\right)\nonumber\\
&=&\frac{2n}{v_{t}^{3}}g^{\al\, 1}\frac{1}{s}\frac{d}{ds}\left(\frac{\erf}{s}\right)\label{lhs}
\end{eqnarray}
For the right-hand side of  (\ref{divd}) we use (\ref{dm2}) to write
\[
-\frac{m}{T}v_{t}s_{\bt}D_{M}^{\al\bt} = -\frac{n}{v_{t}^{2}}s^{\al}(M_{1} - s^{2}M_{2})
\]
where $s^{\al}$ is the normalized contravariant velocity of (\ref{defscontra}).  In quasi-invariant coordinates,
\[
s^{\al} =  v_{t}^{-1}g^{\al1}
\]
so (\ref{m1m2}) gives 
\[
-\frac{m}{T}v_{t}s_{\bt}D_{M}^{\al\bt} =  -\frac{2n}{v_{t}^{3}}g^{\al1}s^{-3}(\erf - s\erf') = \frac{2n}{v_{t}^{3}}g^{\al 1}\frac{1}{s}\frac{d}{ds}\left(\frac{\erf}{s}\right)
\]
in agreement with (\ref{lhs}).

\section{Perturbed Maxwellian case}
\subsection{Maxwellian limit}
When both distributions are Maxwellian, only energy-derivatives survive and (\ref{rmj3}) reduces to 
\begin{equation}
\begin{split}
\frac{C_{Mab}}{\gamma_{ab}} &=  4\pi\frac{m_{a}}{m_{b}} f_{Ma}f_{Mb} + \left(1 - \frac{m_{a}}{m_{b}}\right)2w\frac{dH_{Mb}}{dw}\frac{df_{Ma}}{dw}+\frac{dG_{Mb}}{dw}\frac{df_{Ma}}{dw} \\
&+ 2w^{2}\left(\frac{d^{2}G_{Mb}}{dw^{2}} + \frac{1}{2w}\frac{dG_{Mb}}{dw}\right)\left(\frac{d^{2}f_{Ma}}{dw^{2}} + \frac{1}{2w}\frac{df_{Ma}}{dw}\right)
\end{split}\label{cmm}
\end{equation}
Here
\[
\frac{df_{Ma}}{dw} = -\frac{m_{a}}{T_{a}}f_{Ma}
\]
while from (\ref{gm}) and the definitions of the $M$'s we find that
\begin{eqnarray}
\frac{dG_{Mb}}{dw} &=& \frac{n_{b}}{2v_{tb}}M_{1}, \label{gm1}\\
\frac{d^{2}G_{Mb}}{dw^{2}}&=& -\frac{n_{b}}{2v_{tb}^{3}}M_{2}\label{gm2}
\end{eqnarray}
Similarly (\ref{hm}) yields
\begin{equation}
\label{dhm}
2w\frac{dH_{Mb}}{dw} = \frac{n_{b}}{v_{tb}}s\frac{d}{ds}\left(\frac{\erf}{s}\right)
\end{equation}
Substituting these expressions into (\ref{cmm}) one finds, after some manipulation,
\begin{equation}
\label{cmm2}
C_{Mab} =\bar{\nu}_{ab}\left(1 - \frac{T_{b}}{T_{a}}\right)f_{Ma}\left[\erf'(s)\left(1 + \frac{m_{b}T_{a}}{m_{a}T_{b}}\right) -\frac{\erf(s)}{s}\right]
\end{equation}
where
\begin{equation}
\label{defnb}
\bar{\nu}_{ab} \equiv  \frac{2\gamma_{ab}m_{a}n_{b}}{m_{b}v_{tb}v_{ta}^{2}}
\end{equation}

\subsection{Linear perturbation}
We write
\[
f_{a} = f_{Ma}(1 + \hat{f}_{a})
\]
and neglect terms of order $\hat{f}^{2}$ to obtain the linearized operator
\begin{equation}
\label{defcl}
C^{\ell}_{ab}(\hat{f_{a}},\hat{f_{b} })= C_{ab}(f_{a}, f_{b}) -C_{Mab} +  \mathcal{O}(\hat{f}^{2})
\end{equation}
We begin with the Landau-Boltzmann form, (\ref{c0}), which reduces to
\begin{equation}
C^{\ell}_{ab} = \frac{\gamma_{ab}}{2}\frac{\partial}{\partial v_{\al}}\int d^{3}v\,'f_{Ma}f'_{Mb}U_{\al\bt}\mathcal{C}_{ab\, \bt}
\label{lb1}
\end{equation}
with
\begin{equation}
\label{defcin}
\mathcal{C}_{ab\, \bt} \equiv \frac{\p \hat{f}_{a}}{\p v_{\bt}} - \frac{m_{a}}{m_{b}}\frac{\p \hat{f}'_{b}}{\p v'_{\bt}} - \frac{m_{a}v_{\bt}}{T_{a}}\left(1 - \frac{T_{a}}{T_{b}}\right)(\hat{f}_{a} + \hat{f}'_{b})
\end{equation}
This form evidently simplifies in the like-particle case.  For unlike particles, one usually resorts to mass-ratio approximation, as discussed in Section \ref{sec:massratio}.  Therefore at this point we specialize to the like-particle operator, given by
\begin{equation}
C^{\ell} = \frac{\gamma}{2}\frac{\partial}{\partial v_{\al}}\int d^{3}v\,'f_{M}f'_{M}U_{\al\bt} \left(\frac{\p \hat{f}}{\p v_{\bt}} - \frac{\p \hat{f}'}{\p v'_{\bt}}\right)\label{caal}
\end{equation}
with species labels suppressed.

After writing (\ref{caal}) in covariant form, we integrate by parts and then simplify using (\ref{divd}).  The result is
\begin{equation}\label{cl2}
\begin{split}
C^{\ell}& = \gamma f_{M}\left[4\pi f_{M}\hat{f} +\frac{1}{2}\left(\hat{f}_{,\al,\bt} - \frac{2m}{T}w_{,\al}\hat{f}_{,\bt}\right)D_{M}^{\al\bt}\right.\\
&-\left.\frac{m}{T}H_{1} + \frac{1}{2}\left(\frac{m}{T}\right)^{2}w_{,\al}w_{,\bt}D_{1}^{\al\bt}\right]\end{split}
\end{equation}
Here
\begin{eqnarray}
H_{1}(\hat{f})&\equiv& \int d^{3}v'\, f_{M}'\hat{f}' u^{-1}\label{defh1}\\
D_{1}^{\al\bt}&\equiv& g^{\al\kappa}g^{\bt\lambda}\int d^{3}v' \,f_{M}'\hat{f}' u_{,\kappa,\lambda}\label{defd1}
\end{eqnarray}
The factor of $2$ in the third term of (\ref{cl2}) deserves comment.  It comes from the sum of two contributions, one being the obvious derivative of the Maxwellian. But there is an equal contribution, because of (\ref{divd}), from the derivative of the Maxwellian diffusion tensor. 

Finally we introduce
\begin{equation}
G_{1}(\hat{f}) \equiv  \int d^{3}v' \, f_{M}'\hat{f}' u\label{defg1}
\end{equation}
and recall (\ref{defscontra}) and (\ref{dg2}) in order to express (\ref{cl2}) in the form
\begin{equation}\label{cl3}
\begin{split}
C^{\ell} = &\gamma f_{M}\frac{n}{v_{t}}\left[ \frac{2\erf'}{v_{t}^{2}}\hat{f} +\frac{1}{4}\left(g^{\al\bt}M_{1} -s^{\al}s^{\bt}M_{2}\right) \left(\hat{f}_{,\al,\bt} - \frac{4}{v_{t}}s_{\al}\hat{f}_{,\bt} \right) \right] \\
&+ \gamma f_{M}\frac{m}{T}\left( s^{\al}s^{\bt}G_{1 \,,\al,\bt}-H_{1} \right) 
\end{split}
\end{equation}

\subsection{Linear operator in quasi-invariant coordinates}
We use (\ref{cl2}) to write the linear operator in quasi-invariant coordinates.  The result is simplified using (\ref{m1m2}) and the identity
\[
1 - \frac{3}{2}\lambda B =\sqrt{1 - \lambda B}\frac{\p}{\p\lambda}\lambda\sqrt{1 - \lambda B}
\]
Thus we find
\begin{equation}\label{cl4}\begin{split}
C^{\ell} &=\gamma f_{M}\frac{n}{v_{t}}\left[\frac{M_{1}}{2wB}\left(\sqrt{1 - \lambda B}\frac{\p}{\p \lambda} \lambda \sqrt{1 - \lambda B}\frac{\p \hat{f}}{\p \lambda} + \frac{1}{4\lambda B}\frac{\p^{2}\hat{f}}{\p \gamma^{2}}\right)\right.\\  +& \frac{\erf(s) - s\erf'(s)}{s}\frac{T}{m}\frac{\p^{2}\hat{f}}{\p w^{2}} + \frac{2s\erf'(s) - \erf(s)}{s}\frac{\p \hat{f}}{\p w}\\
+& \left.\frac{m}{T}\erf'(s) \hat{f}\right] + \gamma  f_{M}\frac{m}{T}\left[\left(2w\frac{\p^{2}G_{1}}{\p w^{2}} + \frac{\p G_{1}}{\p w}\right) - H_{1}\right]
\end{split}
\end{equation}
We could also use (\ref{lgh}) and (\ref{del2s}) to write
\begin{eqnarray}
H_{1} & = & \frac{1}{2}\nabla^{2}G_{1}\nonumber\\
& = &\frac{\sqrt{1 - \lambda B}}{wB}\frac{\p}{\p \lambda}\lambda\sqrt{1 - \lambda B}\frac{\p G_{1}}{\p \lambda}\label{del2g}\\&+ &\frac{1}{w^{1/2}}\frac{\p}{\p w}w^{3/2}\frac{\p G_{1}}{\p w} + \frac{1}{4w\lambda B}\frac{\p^{2}G_{1}}{\p \gamma^{2}}\nonumber 
\end{eqnarray}
and thus express the linear operator in terms of $G_{1}$ alone; however the form (\ref{cl4}) is usually most convenient.

\subsection{Linear operator in spherical coordinates}
Returning to (\ref{cl3}), we insert the geometrical coefficients from Subsection \ref{subsec:spher} to obtain the linear operator
\begin{equation}\label{cspher1}\begin{split}
\frac{C^{\ell}}{\gamma_{c} f_{M}} &= \frac{n}{4v_{t}}\left[ M_{1}\nabla^{2}\hat{f} - s^{2}M_{2}\frac{\p^{2}\hat{f}}{\partial v^{2}} - \frac{2m}{T}(M_{1} - s^{2}M_{2})v\frac{\p \hat{f}}{\partial v}\right] \\
&+  4\pi f_{M}\hat{f} + \frac{m}{T}\left(s^{2}\frac{\p^{2}G_{1}}{\partial v^{2}} - H_{1}\right)
\end{split}
\end{equation}
The first term here could be made explicit by means of (\ref{del2spher}).  The $c$-subscript on the coefficient $\gamma_{c}$ distinguishes this constant from the spherical coordinate $\gamma$.

\section{Small mass-ratio approximations}\label{sec:massratio}
\subsection{Scattering of electrons by ions}
For simplicity and concreteness we consider here a plasma with electrons and a single ion species; the relevant small parameter is $m_{e}/m_{i}$.  All small mass-ratio approximations are based on the assumptions that both distributions are nearly Maxwellian, with comparable temperatures.  Thus thermal speeds can be defined for both species, and $v_{te}\gg v_{ti}$.
\subsubsection*{Cartesian coordinates}
To find the approximate form of $C_{ei}$ we can begin with (\ref{c0}), using Cartesian coordinates.  After dropping the term with an explicit $m_{e}/m_{i}$-factor, we have 
\begin{equation}
\label{cei1}
C_{ei }= \frac{\gamma_{ei}}{2} \frac{\p}{\p v_{\al}}\int d^{3}v\,' \,f'_{i}U_{\al\bt}\frac{\p f_{e}}{\p v_{\bt}}
\end{equation}
The sole additional approximation is to write
\[
\int d^{3}v\,' f'_{i}\; U_{\al\bt}(\bv  - \bv') \approx n_{i}U_{\al\bt}(\bv  - \bV_{i})
\]
where $\bV_{i}$ is the mean ion flow, 
\[
n_{i} \bV_{i}\equiv \int d^{3}v\, f_{i}\bv,
\]
and we implicitly neglect terms involving more than one power of $V_{i}$, assuming $V_{i}\ll v_{ti}$.  The nonlinear $C_{ei}$ is therefore
\begin{equation}
\label{cei2}
C_{ei }= \frac{\gamma_{ei}n_{i}}{2} \frac{\p}{\p v_{\al}}U_{\al\bt}(\bv - \bV_{i})\frac{\p f_{e}}{\p v_{\bt}}
\end{equation}

One then finds that 
\[
U_{\al\bt}(\bv - \bV_{i}) =  V_{\al\bt} + V_{\al\bt}^{(1)}
\]
where
\[
V_{\al\bt} = U_{\al\bt}(\bv' \rightarrow 0) = \frac{v^{2}\delta_{\al\bt} - v_{\al}v_{\bt}}{v^{3}}
\]
and
\[
V_{\al\bt}^{(1)} = \frac {v_{\al}V_{i\bt} + v_{\bt}V_{i\al} + \delta_{\al\bt}\bV_{i}\cdot \bv}{v^{3}} - 3 \frac{v_{\al}v_{\bt}\bV_{i}\cdot \bv}{v^{5}}
\]
Notice that (\ref{uu0}) implies
\begin{equation}
\label{vviv}
v_{\al}V_{\al\bt}^{(1)} = V_{i\al}V_{\al\bt}
\end{equation}

The linear version has particular importance.  Using
\[
\frac{\p f_{e}}{\p v_{\bt}} = f_{Me} \left[\frac{\p \hat{f}_{e}}{\p v_{\bt}} -\frac{2v_{\bt}}{v_{te}^{2}}f_{Me} \left(1 + \hat{f}_{e}\right)\right]
\]
an obvious consequence of (\ref{id2}),
\[
\frac{\p V_{\al\bt}}{\p v_{\al}}  = -\frac{2v_{\bt}}{v^{3}}
\]
and (\ref{vviv}), we find
\begin{equation}
\label{cei3}
C^{\ell}_{ei }= \frac{\gamma_{ei}n_{i}}{2}f_{Me} \frac{\p}{\p v_{\al}}\left(V_{\al\bt}\frac{\p \hat{f}_{e}}{\p v_{\bt}} - \frac{4 V_{i\,\al}}{v_{te}^{2}v} \right)
\end{equation}
or, in view of (\ref{gt}),
\begin{eqnarray}
C^{\ell}_{ei }&=& \frac{3\sqrt{\pi}v_{te}^{3}}{8\tau_{ei}}f_{Me} \frac{\p}{\p v_{\al}}\left(V_{\al\bt}\frac{\p \hat{f}_{e}}{\p v_{\bt}} - \frac{4 V_{i\,\al}}{v_{te}^{2}v} \right)\nonumber \\\label{cei3a}
&=&\frac{3\sqrt{\pi}v_{te}^{3}}{8\tau_{ei}}f_{Me} \frac{\p}{\p v_{\al}}V_{\al\bt}\frac{\p \hat{f}_{e}}{\p v_{\bt}} + \frac{3\sqrt{\pi}v_{te} V_{i\,\al}v_{\al}}{2\tau_{ei}v^{3}} f_{Me}
\end{eqnarray}
Here a significant feature is that the operator $C^{\ell}_{ei }$ is not homogeneous in $\hat{f}_{e}$, because of the term involving $V_{i}$. This term represents the contribution from the unperturbed electron distribution, $C_{ei}(f_{Me}, f_{1i})$. In other words the two terms in (\ref{cei3a}) correspond to
\[
C_{ei}^{\ell}(f_{e}, f_{i}) = C_{ei}(f_{1e}, f_{Mi}) + C_{ei}(f_{Me}, f_{1i})  
\]
It is sometimes convenient to denote the homogeneous part of $C_{ei}^{\ell}$ by
\[
 C_{ei}^{h} = C_{ei}^{\ell}\bigg{|}_{V_{i} = 0}
 \]
and thus to write (\ref{cei3a}) as
\begin{equation}
\label{cei3b}
C_{ei}^{\ell} = C_{ei}^{h} + \frac{3\sqrt{\pi}v_{te} V_{i\,\al}v_{\al}}{2\tau_{ei}v^{3}} f_{Me}
\end{equation}

The form (\ref{cei3}) can also be derived from the Galilean invariance of $C_{ei}$, as expressed by (\ref{cinv}).  Under the transformation $\bv \rightarrow \rv  = \bv -\bV_{i}$ the ion distribution
\[
f_{i}(\bv ) \approx f_{Mi}(\bv - \bV_{i})
\]
becomes
\[
\bar{f}_{i}(\rv) = f_{i}(\bv) = f_{i}(\rv + \bV_{i}) = f_{Mi}(\rv),
\]
a Maxwellian at rest.  The electron distribution,
\[
f_{e}(\bv) = f_{Me}(\bv)[1 + \hat{f}_{e}(\bv)]
\]
becomes
\begin{eqnarray}
\bar{f}_{e}(\rv) &=& f_{e}(\bv) = f_{e}(\rv + \bV_{i})\nonumber \\
&=& f_{Me}(\rv + \bV_{i})[1 + \hat{f}_{e}(\rv + \bV_{i})]\nonumber \\
&\approx& f_{Me}(\rv)\left(1 - \frac{m_{e}\rv\cdot \bV_{i}}{T_{e}} \right)[1 + \hat{f}_{e}(\rv)]\nonumber \\
&= & f_{Me}(\rv) \left[1 - \frac{m_{e}\rv\cdot \bV_{i}}{T_{e}} +\hat{f}_{e}(\rv)\right]\label{fed}
\end{eqnarray}
Therefore (\ref{cinv}) implies 
\begin{equation}
\label{galinv}
C_{ei}^{\ell}(\hat{f}_{e}) = C_{ei}^{h}(\hat{f}_{e} - m_{e}\bV_{i}\cdot \bv/T_{e})  
\end{equation}
That (\ref{cei3}) satisfies this relation follows from the identity
\begin{equation}
\label{thm }
C_{ei}^{h}\left(\frac{m_{e}}{T_{e}}\bV\cdot\bv\right) = - \frac{3\sqrt{\pi}v_{te} V_{\al}v_{\al}}{2\tau_{ei}v^{3}} f_{Me}
\end{equation}

We take note here of the approximate Rosenbluth potentials:
\begin{eqnarray}
G_{i} & \approx &  n_{i}v\left(1 - v^{-2}\bV_{i}\cdot \bv\right)\label{gapprox} \\
H_{i} & \approx & (n_{i} /v) \left(1 + v^{-2}\bV_{i}\cdot \bv\right)\label{happrox}
\end{eqnarray}
These relations imply
\[
D_{i\al\bt} = \frac{\p^{2}G_{i}}{\p v_{\al}\p v_{\bt}} \approx n_{i}U_{\al\bt}(\bv - \bV_{i})
\]
and
\[
\frac{\p H_{i}}{\p v_{\bt}} \approx \left(\frac{v_{\bt} - V_{i\bt}}{v^{3}} + 3\frac{\bV_{i}\cdot \bv v_{\bt}}{v^{5}}\right)
\]
In view of (\ref{lgh}), this quantity can be written as
\begin{equation}
\label{hdv}
\frac{\p H_{i}}{\p v_{\bt}} =\frac{n_{i}}{2}\frac{\p}{\p v_{b}} U_{\al\bt}(\bv - \bV_{i})
\end{equation}
It is then straightforward to confirm that (\ref{gapprox}) -- (\ref{hdv}) yield (\ref{cei2}).

\subsubsection*{Quasi-invariant coordinates}
Here, in addition to using quasi-invariant coordinates $(w, \lambda, \gamma)$, we assume that the kinetic equation (``drift-kinetic equation'') involves only the gyrophase average of the distribution:
\[
f(w,\lambda, \gamma) \approx \bar{f}(w, \lambda)
\]
The overbar is henceforth suppressed. For such a distribution, the mean velocity is parallel to the magnetic field:
\[
\bV =  \bb V_{\parallel}
\]
Drift-kinetic theory treats perpendicular flow, which can be comparable to $V_{\parallel}$, by means other than the kinetic equation. Using (\ref{gapprox}) and (\ref{happrox}) we find
\begin{equation}\label{cei4}\begin{split}
C_{ei} = \gamma_{ei}n_{i} &\left[ \frac{\xi}{wB}\left(\frac{1}{\sqrt{2w}} + \frac{V_{\parallel i}\sigma \xi}{2w}\right)\frac{\p}{\p \lambda}\lambda \xi  \frac{\p f_{e}}{\p \lambda}\right. \\
-&\left. \frac{V_{\parallel i} \sigma \xi}{2w} \frac{\p}{\p w}\left( f_{e} + 2\lambda \frac{\p f_{e}}{\p \lambda} \right)\right] \end{split}
\end{equation}
where $\xi \equiv \sqrt{1 - \lambda B}$ as in Subsection \ref{subsec:qic}.  

The linear version of this operator is
\begin{equation}\label{cei5}
C_{ei}^{\ell} = \gamma_{ei}n_{i} f_{Me} \frac{\xi}{2w^{2}}\left(\frac{\sqrt{2w}}{B}\frac{\p}{\p \lambda}\lambda \xi  \frac{\p \hat{f}_{e}}{\p \lambda} + \frac{m_{e}V_{\parallel i}\sigma w}{T_{e}}\right)
\end{equation}
Although $C_{ei}$ is a rotationally symmetric operator, the right-hand side of (\ref{cei5}) is not a symmetric function of velocity, as manifested by the factor $\sigma$ that occurs with $V_{\parallel i}$.  Ion parallel flow breaks the symmetry: $\bar{f}_{i}(\rv) = f_{i}(\bv) \neq f_{i}(\rv)$.  This circumstance was anticipated at the end of subsection \ref{subsec:inv}.

The coefficient $\gamma_{ei}$ contains a factor $e_{i}^{2} = Z^{2}e^{2}$ and therefore can exceed $\gamma_{ee}$ for multiply charged ion species.  In other respects however, we see that 
\[
C_{ei} \sim C_{ee}
\]
That is, electrons are scattered by ions at a rate comparable to their self-scattering.

\subsection{Scattering of ions by electrons}
A convenient starting point for $C_{ie}$ is (\ref{rmj3}); using Cartesian coordinates for simplicity we have 
\begin{equation}
\label{cie1}\begin{split}
C_{ie} =&   \frac{\gamma_{ie}}{2}\left[8\pi \frac{m_{i}}{m_{e}}f_{i}f_{e} + \frac{\p^{2}f_{i}}{\p v_{\al}\p v_{\bt}}\frac{\p^{2}G_{e}}{\p v_{\al}\p v_{\bt}}\right. \\ &+ \left. 2\left(1 - \frac{m_{i}}{m_{e}}\right)\frac{\p f_{i}}{\p v_{\bt}}\frac{\p H_{e}}{\p v_{\bt}}\right]\end{split}
\end{equation}
In the first two terms on the right-hand side we need retain only the contributions from the electron Maxwellian; the non-Maxwellian electron distribution $\hat{f}_{e}$ has no special qualitative significance and offers only a small correction. Thus   
\[ 
\frac{m_{i}}{m_{e}}f_{e}f_{i}\approx  \frac{m_{i}}{m_{e}}\frac{n_{e}}{v_{te}^{3}}f_{i} 
\]
since $v/v_{te} \sim v_{ti}/v_{te}$ is neglected. For the second term, involving $G_{e}\approx G_{Me}$, we use (\ref{dg2}), (\ref{dm2}) and (\ref{asym2}) to find
\[
\frac{\p^{2}G_{e}}{\p v_{\al}\p v_{\bt}}\approx \delta_{\al\bt}\frac{4n_{e}}{3\sqrt{\pi}v_{te}}
\]

The last term in (\ref{cie1}), involving $H_{e}$, is slightly more complicated.   The point is that a straightforward mass-ratio expansion of $H_{e}$ gives
\begin{equation}
\label{hex}
\frac{\p H_{e}}{\p v_{\bt}} = \int d^{3}v  \,' f_{e}' |v'|^{-3}(v'_{\bt} + \cdots)
\end{equation}
Here the leading term is of order $n_{e}v_{te}^{-2}\hat{f}_{e}$: it is relatively large with respect to the mass ratio, but can contribute only in the case of a non-isotropic perturbation to the electron distribution.  Thus the perturbation $\hat{f}_{e}$ does have special qualitative significance in this case, and we must retain both $H_{Me}$ and $H_{1e}$.  For $H_{Me}$ we use (\ref{hm}) to find
\[
\frac{\p H_{Me}}{\p v_{\bt}} \approx -\frac{4n_{e}}{3\sqrt{\pi}v_{te}^{3}} v_{\bt}
\]
The lowest-order contribution to the correction is, as noted above,
\begin{equation}
\label{h1e}
\frac{\p H_{1e}}{\p v_{\bt}} = \int d^{3}v\,'\, f_{e}' \frac{v'_{\bt}}{v'^{3}}
\end{equation}
In Section \ref{sec:mom} we show that this quantity is related to the collisional friction force $\bF_{ei}$:
\begin{equation}
\label{phf}
\frac{\p H_{1e}}{\p v_{\bt}} = \frac{n_{e}V_{i\bt}}{\gamma_{ei}\tau_{ei}n_{i}} - \frac{F_{ei\, \bt}}{\gamma_{ei}n_{i}m_{e}}
\end{equation}

After combining these results and recalling (\ref{gt}), we obtain the operator
\begin{eqnarray}
C_{ie} &=& \frac{m_{e}n_{e}}{m_{i}n_{i}\tau_{ei}}\left[3f_{i} + \frac{T_{e}}{m_{i}}\frac{\p^{2}f_{i}}{\p v_{\al}\p v_{\bt}} + (v_{\bt} - V_{i\bt})\frac{\p f_{i}}{\p v_{\bt}}\right]+\frac{F_{ei\, \bt}}{m_{i}n_{i}}\frac{\p f_{i}}{\p v_{\bt}}\nonumber \\
&=& \frac{m_{e}n_{e}}{m_{i}n_{i}\tau_{ei}}\frac{\p }{\p v_{\bt}}\left[(v_{\bt} - V_{i\bt})f_{i} + \frac{T_{e}}{m_{i}}\frac{\p f_{i}}{\p v_{\bt}} \right]+\frac{F_{ei\, \bt}}{m_{i}n_{i}}\frac{\p f_{i}}{\p v_{\bt}}\label{cie2}
\end{eqnarray}

As a check on (\ref{cie2}), we consider the strictly Maxwellian limit, with $V_{i} = 0 =F_{ei}$:
\begin{eqnarray}
C_{Mie} &=& \frac{m_{e}n_{e}}{m_{i}n_{i}\tau_{ei}}\frac{\p }{\p v_{\al}}\left(v_{\al}f_{Mi} + \frac{T_{e}}{m_{i}}\frac{\p f_{Mi}}{\p v_{\al}} \right)\nonumber \\
&=&\frac{m_{e}n_{e}}{m_{i}n_{i}\tau_{ei}}\left( 1 - \frac{T_{e}}{T_{i}}\right)f_{Mi}\left(3 - \frac{2v^{2}}{v_{ti}^{2}}\right)\label{cm4}
\end{eqnarray}
To compare this result with the corresponding expansion of (\ref{cmm2}), we write the latter as
\begin{equation}
\label{cm5}
C_{Mie} = \frac{2\gamma_{ie}m_{i}n_{e}}{m_{e}v_{te}^{3}}\left(1 - \frac{T_{e}}{T_{i}}\right)\mathcal{C}(s)
\end{equation}
where
\[
\mathcal{C}(s) =  \left(1 + \frac{m_{i}T_{e}}{m_{e}T_{i}}\right)\erf'(s) - \frac{m_{i}T_{e}}{m_{e}T_{i}}\frac{\erf(s)}{s}
\]
and expand $\mathcal{C}$ for $s\ll1$:
\begin{eqnarray*}
\mathcal{C} &=& \frac{2}{\sqrt{\pi}}\left[1 -\frac{m_{i}T_{e}}{m_{e}T_{i}}\left(\frac{2}{3}s^{2} + \mathcal{O}(s^{4})\right)\right]\\
& =& \frac{2}{3\sqrt{\pi}}\left[3 -2\frac{v^{2}}{v_{ti}^{2}}\left(1 + \mathcal{O}(s^{2})\right)\right]
\end{eqnarray*}
In view of (\ref{gt}), substitution of this result into (\ref{cm5}) reproduces (\ref{cm4}).

Returning to (\ref{cie2}), we use (\ref{phf}) and (\ref{gt}) to rewrite it as
\begin{equation}
\label{cie3}
C_{ie} = \frac{m_{e}}{m_{i}\tau_{ei}}\left(\frac{T_{e}}{m_{i}}\frac{\p^{2}f_{i}}{\p v_{\bt}\p v_{\bt}} + \frac{\p}{v_{\bt}}(v_{\bt}f_{i}) - \frac{3\sqrt{\pi}v_{te}^{3}}{4n}\frac{\p H_{1e}}{\p v_{\bt}}\frac{\p f_{i}}{\p v_{\bt}}\right)
\end{equation}
Now the first two terms on the right here are estimated by
\[
\frac{m_{e}}{m_{i}\tau_{ei}}\left(\frac{T_{e}}{m_{i}}\frac{\p^{2}f_{i}}{\p v_{\bt}\p v_{\bt}} + \frac{\p}{v_{\bt}}(v_{\bt}f_{i})\right) \cong \frac{m_{e}}{m_{i}\tau_{ei}}f_{i}
\]
For the third term we use (\ref{h1e}) to get the nominal estimate
\[
\frac{\p H_{1e}}{\p v_{\bt}}\cong \frac{n}{v_{te}^{2}}
\]
whence
\[
\frac{3\sqrt{\pi}v_{te}^{3}}{4n} \frac{\p H_{1e}}{v_{\bt}}\frac{\p f_{i}}{\p v_{\bt}}\cong \frac{m_{e}}{m_{i}\tau_{ei}}f_{i}\frac{v_{te}}{v_{ti}}
\]
In other words, with the nominal ordering $C_{ie}\approx C_{ie}^{*}$ where
\begin{equation}
\label{cie4}
C_{ie}^{*} = -\frac{3\sqrt{\pi}v_{te}^{3}m_{e}}{4nm_{i}\tau_{ei}}  \frac{\p H_{1e}}{v_{\bt}}\frac{\p f_{i}}{\p v_{\bt}}  = -\gamma_{ei}\frac{m_{e}}{m_{i}}\frac{\p f_{i}}{\p v_{\bt}}\int \dif^{3}v' f'_{e}\frac{v'_{\bt}}{v^{'3}}
\end{equation}
We next use (\ref{phf}) to write this approximation in term of the friction force; again a nominal ordering has
\[
\frac{nm_{e}}{\tau_{ei}}V_{i\bt} \ll F_{ei\bt}
\]
in which case
\begin{equation}
\label{cie5}
C_{ie}^{*} = \frac{1}{nm_{i}} F_{ei\bt}  \frac{\p f_{i}}{\p v_{\bt}}
\end{equation}
Thus in the nominal ordering (which may not always apply) electron scattering of ions amounts to no more than electron friction adding to the total force on the ions.

To linearize the operator we use (\ref{cie4}), noticing that the integral vanishes for Maxwellian $f_{e} = f_{Me}$.  Therefore
\begin{equation}
\label{ciel}
C_{ie}^{\ell} = \gamma_{ei}\frac{m_{e}}{T_{i}}f_{Mi}v_{\bt}\int \dif^{3}v' f'_{Me}\hat{f}_{e}\frac{v'_{\bt}}{v^{'3}}
\end{equation}

\section{Moments of the operator} \label{sec:mom}
The zeroth moment of the Coulomb collision operator is easily seen to vanish:
\[
\int d^{3}v\, C_{ab} = 0,
\]
for any physically reasonable distributions.  The other moments of general interest are the collisional momentum exchange, or ``friction force,''
\begin{equation}
\label{deffric}
\bF_{ab}\equiv \int d^{3}v\, m_{a}\bv C_{ab}
\end{equation}
and the collisional energy exchange
\begin{equation}
\label{defqex}
W_{ab} \equiv \int d^{3}v\, \frac{1}{2}m_{a}v^{2} C_{ab}
\end{equation}
We consider these moments in order.
\subsection{Friction force}
It is convenient to write the collision operator in RMJ form, using (\ref{rmj}).  After substitution into (\ref{deffric}) and partial integration, we find
\begin{equation}
\label{fric1}
\bF_{ab} = -m_{a}\gamma_{ab}\left(1 +\frac{m_{a}}{m_{b}}\right)\int d^{3}v\, f_{a}\frac{\p H_{b}}{\p \bv}
\end{equation}

To proceed further, we specialize to $\bF_{ei}$ and use a mass-ratio approximation. It is awkward to apply the mass-ratio expansion directly to (\ref{fric1}), because of singularities that appear in the expansion of $|\bv - \bv'|^{-3}$, so we use (\ref{cei3}).  (The following argument has appeared previously\cite{meiss92}; see also\cite{braginskii}.)  After integration by parts we have
\[
F_{ei\, \gamma} = -\frac{\gamma_{ei}m_{e}n_{i}}{2}\int d^{3}\,v\left(V_{\al\bt}\frac{\p \hat{f}_{e}}{\p v_{\al}} - 4\frac{V_{i\,\bt}}{v_{te}^{2}v}\right) \frac{\p}{\p v_{\bt}}v_{\gamma} f_{Me}
\]
But
\[
\frac{\p}{\p v_{\bt}}v_{\gamma} f_{Me} = f_{Me}\left(\delta_{\bt\gamma} - 2\frac{v_{\bt}v_{\gamma}}{v_{te}^{2}}\right)
\]
and
\[
\int d^{3}v\, f_{Me}\frac{1}{v}\left(\delta_{\bt\gamma} - 2\frac{v_{\bt}v_{\gamma}}{v_{te}^{2}}\right) = \frac{2\sqrt{\pi}}{3}\frac{n_{e}}{v_{te}}\delta_{\bt\gamma}
\]
Therefore, in view of  (\ref{gt}) and (\ref{uu0}),
\begin{equation}
\label{fei1}
F_{ei\, \gamma} = -\frac{\gamma_{ei}m_{e}n_{i}}{2}\int d^{3}v \,f_{Me}V_{\al\gamma}\frac{\p \hat{f}_{e}}{\p v_{\al}} + \frac{m_{e}n_{e}}{\tau_{ei}}V_{i\,\gamma}
\end{equation}
Another integration by parts, with (\ref{id2}), gives an equally useful form:
\begin{equation}
\label{fei2}
\bF_{ei} = -\gamma_{ei}m_{e}n_{i}\int d^{3}v \, f_{e}\frac{\bv}{v^{3}}+ \frac{m_{e}n_{e}}{\tau_{ei}}\bV_{i}
\end{equation}
Comparing (\ref{fric1}) to (\ref{fei2}) confirms (\ref{phf}).  The appearance of $\tau_{ei}$ with unit coefficient reflects the historical definition of the collision time (``momentum exchange time'').

\subsection{Energy exchange}
We compute the energy exchange assuming all species have Maxwellian distributions in a common rest frame.  The simplest calculation uses (\ref{cm4}), but it is somewhat instructive to eschew mass-ratio approximation and use (\ref{cmm2}). Thus we have
\begin{eqnarray}
W_{ab} &=&4\pi \frac{m_{a}}{2}v_{tb}^{5} \int  ds\, s^{4}C_{Mab}\nonumber\\
&=& -2\frac{\gamma_{ab}n_{a}n_{b}m_{a}^{2}}{v_{ta}m_{b}}\left(\frac{v_{tb}}{v_{ta}}\right)^{4}  \left(1 - \frac{T_{b}}{T_{a}}\right)\nonumber\\
&\times & \int_{0}^{\infty} ds \, s^{4}\erf'(\beta s)\left[\frac{\erf(s)}{s} - \left(1 + \frac{m_{b}T_{a}}{m_{a}T_{b}}\right)\erf'(s)\right]\label{wab0}
\end{eqnarray}
Here
\[
\beta \equiv \frac{v_{tb}}{v_{ta}}
\]
so that, in particular,
\[
f_{Ma} = \frac{n_{a}}{2\pi v_{ta}^{3}}\erf'(\bt s)
\]
and furthermore,
\[
1 + \frac{m_{b}T_{a}}{m_{a}T_{b}} = \frac{1 + \bt^{2}}{\bt^{2}}
\]
It is not hard to show, by partial integration, that
\[
\int_{0}^{\infty} ds \, s^{3}\erf'(\beta s)  \erf(s)  = \frac{2}{\sqrt{\pi}}\frac{2 + 3\bt^{2}}{4\bt^{4}(1 + \bt^{2})^{3/2}}
\]
and that
\[
\int_{0}^{\infty} ds\,  s^{4}\erf'(\beta s)  \erf'(s)  = \frac{2}{\sqrt{\pi}}\frac{3}{(1 + \bt^{2})^{5/2}}
\]
After substitution in (\ref{wab0}) we find
\[
W_{ab} = -4\gamma_{ab}n_{a}n_{b}\frac{m_{a}\bt^{4}}{m_{b}v_{ta}^{3}}(T_{a} - T_{b})\mathcal{W}
\]
where $\mathcal{W}$ represents the square-bracketted quantity in (\ref{wab0}), coming from the various integrals:
\begin{eqnarray}
\mathcal{W} &=& \frac{2}{\sqrt{\pi}}\left[\frac{2 + 3\bt^{2}}{4\bt^{4}(1 + \bt^{2})^{3/2}} - \frac{1 + \bt^{2}}{\bt^{2}}\frac{3}{4(1 + \bt^{2})^{5/2}}\right]\nonumber\\
&=&\frac{2}{\sqrt{\pi}}\frac{ (2 + 3\bt^{2} - 3\bt^{2})}{4\bt^{4}(1 + \bt^{2})^{3/2}}\label{w3}
\end{eqnarray}
The sole noteworthy feature of the calculation is the cancellation, in (\ref{w3}), of the two $\bt^{2}$-terms in the numerator; since $\bt^{2} \sim m_{a}/m_{b}$, the result is to insert a perhaps unanticipated factor $m_{b}/m_{a}$ into $W_{ab}$.  We have
\begin{equation}
\label{wabfin}
W_{ab}= -4\gamma_{ab}n_{a}n_{b}\frac{m_{a}}{m_{b}}\frac{T_{a} - T_{b}}{(v_{ta}^{2}+v_{tb}^{2})^{3/2}}
\end{equation}
which pertains for arbitrary mass ratio.  The small mass-ratio limit of $W_{ei}$ coincides with the appropriate moment of (\ref{cm4}).  Note that the right-hand side of (\ref{wabfin}) is antisymmetric under species interchange,
\[
W_{ab} = -W_{ba}
\]
as required by energy conservation.

\section{Quadratic entropy production}
\subsection{Bilinear form}
It can be shown\cite{framework} that the collisional rate of change of Boltzmann entropy density for a single species,
\[
s_{a} \equiv - \int \dif^{3}\, v f_{a}\log f_{a}
\]
is given, through second order in the normalized perturbation $\hat{f}_{a}$, by
\[
\Theta_{a} = -\int\dif^{3}v\,\hat{f}_{a}C_{a}^{\ell}(f)
\]
That is,
\begin{equation}
\label{deftheta0}
\Theta_{a} = \sum_{b}\Theta_{ab}
\end{equation}
with
\begin{equation}
\label{deftheta1}
\Theta_{ab}  \equiv - \int\dif^{3}v\,\hat{f}_{a}C_{ab}^{\ell}(f_{a}, f_{b})
\end{equation}
Thus the total plasma entropy production is
\begin{equation}
\label{tottheta}
\Theta = \sum_{a}\Theta_{a} = \sum_{a,b}\Theta_{ab}
\end{equation}
and it is natural to define the bilinear form
\begin{equation}
\label{defphiab}
\Phi_{ab}[f,g]  =  - \int\dif^{3}v\,\hat{f}_{a}C_{ab}^{\ell}(g_{a}, g_{b}) 
\end{equation}
where the functions $f_{a}$ and $g_{a,b}$ are arbitrary integrable functions---not necessarily solutions to any kinetic equation.  When evaluated on the physical distributions, $\Phi_{ab}$ measures the entropy generated in species $a$ due to collisions with species $b$.  Similarly the form
\begin{equation}
\label{ftot}
\Phi[f,g] = \sum_{a,b}\Phi_{ab}
\end{equation}
measures the total entropy production, when it is evaluated on the physical distributions.

Note that three functions appear in $\Phi_{ab}[f,g]$, while $\Phi[f,g]$ is a functional of four functions.  Thus the $f = (f_{a},f_{b})$ and $g = (g_{a}, g_{b})$ on the left-hand sides of (\ref{defphiab}) and (\ref{ftot}) each represent two-component quantities.

We next show that $\Phi[f,g]$ is symmetric under interchange $f \leftrightarrow g$. This symmetry is usually referred to as a \textit{self-adjointness} property.
\subsection{Equal species temperatures}
Here we suppose that $T_{a} = T_{b}$ for all species pairs $(a,b)$.  Then (\ref{lb1}) implies, after partial integration,
\[
\Phi_{ab}[f,g] = \frac{\gamma_{ab}}{2} \int \dif^{3}v\dif^{3}v'\,f_{Ma}f'_{Mb}U_{\al\bt} \frac{\p \hat{f}_{a}}{\p v_{\al}}\left(\frac{\p \hat{g}_{a}}{\p v_{\bt}} - \frac{m_{a}}{m_{b}}\frac{\p \hat{g}'_{b}}{\p v'_{\bt}}\right)
\]
We introduce the constants
\[
c_{ab} \equiv m_{a}^{2}\gamma_{ab}
\]
which satisfy $c_{ab} = c_{ba}$.  Minor rearrangement provides
\begin{equation}
\label{fab}
\begin{split}
\Phi_{ab}[f,g] &= \frac{c_{ab}}{2}\int \dif^{3}v\dif^{3}v'\,f_{Ma}f'_{Mb}U_{\al\bt}\frac{1}{m_{a}}\frac{\p \hat{f}_{a}}{\p v_{\al}}\\
&\times \left(\frac{1}{m_{a}}\frac{\p \hat{g}_{a}}{\p v_{\bt}} - \frac{1}{m_{b}}\frac{\p \hat{g}'_{b}}{\p v'_{\bt}}\right)
\end{split}
\end{equation}
Next we symmetrize $a\leftrightarrow b, \bv\leftrightarrow \bv'$ simultaneously (thus preserving $f_{Ma}f'_{Mb}$) to obtain
\begin{equation}
\label{fba}
\begin{split}
\Phi_{ba}[f,g] &= \frac{c_{ab}}{2}\int \dif^{3}v\dif^{3}v'\,f_{Ma}f'_{Mb}U_{\al\bt}\frac{1}{m_{b}}\frac{\p \hat{f}'_{b}}{\p v'_{\al}}\\
&\times \left(\frac{1}{m_{b}}\frac{\p \hat{g}'_{b}}{\p v'_{\bt}} - \frac{1}{m_{a}}\frac{\p \hat{g}_{a}}{\p v_{\bt}}\right)
\end{split}
\end{equation}
Observing that the parenthesized factors on the right-hand sides of (\ref{fab}) and (\ref{fba}) differ only by sign, we add to conclude that
\begin{equation}
\label{ffin}
\begin{split}
\Phi_{ab}[f,g] &+ \Phi_{ba}[f,g] =  \frac{c_{ab}}{2}\int \dif^{3}v\dif^{3}v'\,f_{Ma}f'_{Mb}U_{\al\bt}\\
&\times \left(\frac{1}{m_{a}}\frac{\p \hat{f}_{a}}{\p v_{\al}} - \frac{1}{m_{b}}\frac{\p \hat{f}'_{b}}{\p v'_{\al}}\right)  \left(\frac{1}{m_{a}}\frac{\p \hat{g}_{a}}{\p v_{\bt}} - \frac{1}{m_{b}}\frac{\p \hat{g}'_{b}}{\p v'_{\bt}}\right)
\end{split}
\end{equation}
Now the symmetry $U_{\al\bt} = U_{\bt\al}$ clearly implies
\begin{equation}
\label{fabba}
\Phi_{ab}[f,g] + \Phi_{ba}[f,g] = \Phi_{ab}[g,f] + \Phi_{ba}[g,f]
\end{equation}
whence
\begin{equation}
\label{sa0}
\Phi_{aa}[f,g] = \Phi_{aa}[g,f] 
\end{equation}
and
\begin{equation}
\label{sa1}
\Phi[f,g] = \Phi[g,f]
\end{equation}
Thus the total bilinear form is self-adjoint in the equal-temperature case.

It is also clear from (\ref{ffin}) that the entropy production $\Theta = \Phi[f,f]$ cannot be negative, for any integrable distributions $f_{a,b}$:
\begin{equation}
\label{pos}
\Phi[f,f] \geq 0
\end{equation}
The point is that the tensor $U_{\al\bt}$ is positive in the sense that
\[
\xi_{\al}U_{\al\bt}\xi_{\bt} = u^{-3}[u^{2}\xi^{2} - (\bm{\xi}\cdot \bm{u})^{2}] \geq 0
\]  
for any vector function $\xi(\bv, \bv')$.
\subsection{Unequal temperatures}
Here we restrict attention to the case of a single ion species and use the mass-ratio approximations developed in Section \ref{sec:massratio}.  The relevant linearized operators are given by (\ref{cei3b}),
\[
C_{ei}^{\ell} = C_{ei}^{h} + \gamma_{ei}\frac{m_{e}}{T_{e}}f_{Me} V_{i\,\bt}\frac{v_{\al}}{v^{3}} 
\]
and (\ref{ciel}),
\[
C_{ie}^{\ell} = \gamma_{ei}\frac{m_{e}}{T_{i}}f_{Mi}v_{\bt}\int \dif^{3}v' f'_{Me}\hat{f}_{e}\frac{v'_{\bt}}{v^{'3}}
\]
We observe first that the homogeneous operator $C_{ei}^{h}$, which does not explicitly depend upon $f_{i}$, is self-adjoint in itself; that is, if
\[
\Phi^{h}_{ei}[f,g] \equiv - \int\dif^{3}v\,\hat{f}_{e}C_{ei}^{h}(g_{e}) 
\]
then
\[
\Phi^{h}_{ei}[f,g] = \Phi^{h}_{ei}[g,f]
\]
We therefore restrict attention to
\[
\Phi_{ei}^{*} \equiv   -\gamma_{ei}\frac{m_{e}}{T_{e}}V_{i\,\bt}[g_{i}]\int\dif^{3}v\,f_{Me}\hat{f}_{e} \frac{v_{\bt}}{v^{3}}
\]
or, more explicitly,
\begin{equation}
\label{fei*}
\Phi_{ei}^{*} \equiv  - \gamma_{ei}\frac{m_{e}}{T_{e}}\int \dif^{3}v'\,f'_{Mi}\hat{g}'_{i}v'_{\bt}\int\dif^{3}v\,\,f_{Me}\hat{f}_{e} \frac{v_{\bt}}{v^{3}}
\end{equation}
Comparing this quantity to
\[
\Phi_{ie} = - \gamma_{ei}\frac{m_{e}}{T_{i}}\int\dif^{3}v\, f_{Mi}\hat{f}_{i}v_{\bt}\int \dif^{3}v' f'_{Me}\hat{g}_{e}\frac{v'_{\bt}}{v^{'3}}
\]
we see that the quantity $T_{e}\Phi_{ei}^{*} + T_{i}\Phi_{ie}$ is self-adjoint, and therefore
\begin{equation}
\label{sa3}
T_{e}\Phi_{ei}[f,g] + T_{i}\Phi_{ie}[f,g] = T_{e}\Phi_{ei}[g,f] + T_{i}\Phi_{ie}[g,f]
\end{equation}
In other words, for unequal-species temperatures the bilinear form $\Phi$ is no longer self-adjoint.  However, the small mass-ratio approximation provides a distinct form,
\begin{equation}
\label{sa5}
\Xi[f,g] \equiv T_{e}\Phi_{ei}[f,g] + T_{i}\Phi_{ie}[f,g]
\end{equation}
that is self-adjoint.

\section{Expansion in spherical harmonics}
\subsection{Separated distribution}
We consider the linearized, like-particle collision operator in spherical coordinates.  We begin by writing (\ref{cspher1}) in a slightly more convenient form, by introducing the differential rotation operator $\mathcal{R}$, in terms of which
\[
\nabla^{2}\hat{f} = \frac{1}{s^{2}}\frac{\p}{\p s}s^{2}\frac{\p \hat{f}}{\p s}+ \frac{1}{s^{2}}\mathcal{R}^{2}\hat{f}
\]
and by defining the normalized potentials
\begin{eqnarray}
\hat{G}(\hat{f}) &=& \frac{\sqrt{\pi}G_{1}(\hat{f})}{2nv_{t}} \label{defghat}\\
\hat{H}(\hat{f}) &=& \frac{\sqrt{\pi}v_{t }H_{1}(\hat{f})}{2n}\label{defhhat}
\end{eqnarray}
where the right-hand sides are given by (\ref{defh1}) and (\ref{defg1}).  Then  (\ref{cspher1}) becomes
\begin{equation}\label{cl2a}
\begin{split}
C^{\ell}(\hat{f})  &=\frac{\gamma_{c} n}{4v_{t}^{3}}f_{M}\left[(M_{1}-s^{2}M_{2})\left(\frac{\p^{2}\hat{f}}{\p s^{2}} - 4s\frac{\p\hat{f}}{\p s}\right) \right. \\ & \left. +\frac{M_{1}}{s^{2}}\left(2s \frac{\p \hat{f}}{\p s} + \mathcal{R}^{2}\hat{f}\right) + 8\erf'(s)\hat{f} + \frac{16}{\sqrt{\pi}}\left(s^{2}\frac{\p^{2}\hat{G}}{\p s^{2}} - \hat{H}\right) \right]
\end{split}
\end{equation}

Roational symmetry suggests expansion of the distribution in terms of  spherical harmonics $Y_{\ell m}(\theta, \gamma)$:
\[
\hat{f} = \sum_{\ell,m}F_{\ell m}(s)Y_{\ell m}(\theta, \gamma)
\]
where the functions $F_{\ell m}(s)$ are projections of $\hat{f}$. Using
\[
\mathcal{R}^{2}Y_{\ell m} = -\ell(\ell+1)Y_{\ell m}
\]
and defining the functions $G_{\ell m}(s)$ and $H_{\ell m}(s)$ by
\[
\hat{G}(F_{\ell m}Y_{\ell m}) =G_{\ell m}(s)Y_{\ell m}, \,\,\hat{H}(F_{\ell m}Y_{\ell m}) = H_{\ell m}(s)Y_{\ell m}
\]
we obtain the expression
\begin{equation}\label{cl5}
\begin{split}
C^{\ell}(F_{\ell m}Y_{\ell m})  &=\frac{\gamma_{c} n}{4v_{t}^{3}}f_{M}Y_{\ell m}\left[(M_{1}-s^{2}M_{2})(F'' - 4sF')+2s^{-1}M_{1} F' \right. \\ & \left.  -\ell(\ell +1)s^{-2} M_{1} F + 8\erf'F +  \frac{16}{\sqrt{\pi}}(s^{2}G'' -H) \right]
\end{split}
\end{equation}
Here we have suppressed the subscripts on $F$, $G$ and $H$.
\subsection{Energy variable}
We denote the bracketted expression in (\ref{cl5}) by $(2/\sqrt{\pi})\Psi_{\ell m}$:
\[
C^{\ell}(F_{\ell m}Y_{\ell m}) =\frac{\gamma_{c} n}{2\sqrt{\pi}v_{t}^{3}}f_{M}Y_{\ell m}\Psi_{\ell m}
\]
It is useful to express $\Psi_{\ell m}$ in terms of the variable
\[
x = s^{2}
\]
measuring kinetic energy.  We introduce 
\begin{eqnarray}
\Phi_{1}(x)& = &(\sqrt{\pi}/2)[M_{1}(\sqrt{x}) - xM_{2}(\sqrt{x})] \nonumber\\
&=& x^{-3/2}[\gamma(1/2,x) - 2x^{1/2}\me^{-x}]
\end{eqnarray}
and 
\begin{eqnarray}
\Phi_{2}(x) &=& (\sqrt{\pi}/2)M_{1}(\sqrt{x})\nonumber \\
&=& x^{-3/2}\left[x^{1/2}\me^{-x}+ \left(x - \frac{1}{2}\right) \gamma(1/2, x)\right]
\end{eqnarray}
Here $\gamma$ is the incomplete gamma-function:
\begin{equation}
\label{defgam}
\gamma(\al, x) \equiv \int_{0}^{x} \dif t\, t^{\al - 1}\me^{-t}
\end{equation}
Recall
\[
\gamma(1/2,x) = \sqrt{\pi}\erf(s),\,\,\, e^{-x} = \frac{\sqrt{\pi}}{2}\erf'(x)
\]
Noticing that
\[
d/ds = 2\sqrt{x}d/dx,\;\;(d/ds)^{2} = 4x(d/dx)^{2} + 2d/dx
\]
we find
\begin{equation}
\label{psi}
\begin{split}
\Psi_{\ell m}(x) &= 2x\Phi_{1}(x)F_{xx}+ 2[(1-4x)\Phi_{1} + 2\Phi_{2}(x)]F_{x} \\ &+ [8\me^{-x}- \ell(\ell+1)x^{-1}\Phi_{2}(x)]F + 8(4x^{2}G_{xx} + 2xG_{x} - H)\end{split}
\end{equation}
where the $x$-subscripts indicate derivatives. 

We next express the function $H_{\ell m}$ more explicitly, starting with (\ref{defh1}) and (\ref{defhhat}).  The familiar identity
\begin{equation}
\label{1overu}
\frac{1}{u} =\frac{1}{v_{>}}\sum_{\ell}P_{\ell}(\cos \phi)\left(\frac{v_{<}}{v_{>}}\right)^{\ell}
\end{equation}
where $v_{<}$ and $v_{>}$ are respectively the smaller and larger of the variables $v$ and $v'$, can be combined with 
\begin{equation}
\label{pyy}
P_{\ell}(\cos \phi) = \frac{4\pi}{2\ell + 1}\sum_{m  = -\ell}^{\ell}Y^{*}_{\ell m}(\theta', \gamma')Y_{\ell m}(\theta, \gamma)
\end{equation}
and the orthogonality condition
\[
\int_{0}^{2\pi}\dif \gamma \int_{0}^{\pi}\dif \theta \sin \theta \, Y^{*}_{\ell m}(\theta, \gamma)Y_{\ell' m'}(\theta, \gamma) = \delta_{\ell' \ell}\delta_{m'm}
\]
to provide the expression 
\[
H_{\ell m} = \frac{2}{2\ell + 1}\int_{0}^{\infty}\dif s' \,s'^{2}\frac{s_{<}^{\ell}}{s_{>}^{\ell+1}}\me^{-s'^{2}}F_{\ell m}(s')
\]
or
\begin{equation}
\label{hf3}
H_{\ell m} = \frac{1}{2\ell+1}\int_{0}^{\infty}\dif x' x'^{1/2}\frac{x_{<}^{\ell/2}}{x_{>}^{(\ell+1)/2}}\me^{-x'}F_{\ell m}(x')
\end{equation}
Then the relation (\ref{gib}) can be used to show that
\begin{equation}
\label{gh3}
G_{\ell m} = -\frac{1}{2\pi (2\ell+1)}\int_{0}^{\infty}\dif x' x'^{1/2}\frac{x_{<}^{\ell/2}}{x_{>}^{(\ell+1)/2}}\me^{-x'}H_{\ell m}(x')
\end{equation}
However we find this expression for $G_{\ell m}$ less useful than one that will be derived presently.

We express (\ref{hf3}) as
\[
H_{\ell m} =  \mathcal{L}[F_{\ell m}]
\]
where the operator 
\[
\mathcal{L}[f](x) \equiv \frac{1}{2\ell+1}\int_{0}^{\infty}\dif x'\, x'^{1/2}\frac{x_{<}^{\ell/2}}{x_{>}^{(\ell+1)/2}}\me^{-x'}f(x')
\]
has the more convenient form
\begin{equation}
\label{defL}
\begin{split}
\mathcal{L}[f](x)&=\frac{1}{2\ell+1}\left\{\int_{0}^{\infty}\dif x'\left(\frac{x'}{x}\right)^{(\ell+1)/2}\me^{-x'}f(x')\right. \\
& \left. +\int_{x}^{\infty}\dif x'\left[\left(\frac{x}{x'}\right)^{\ell/2} -\left(\frac{x'}{x}\right)^{(\ell+1)/2}\right]\me^{-x'} f(x')\right\} \end{split}
\end{equation}

\subsection{Laguerre projections}

It is natural to express a smooth distribution in terms of Laguerre polynomials, 
\begin{equation}
\label{deflp}
L_{p}^{\al} \equiv \frac{1}{p!}\me^{x}x^{-\al}\frac{d^{p}}{dx^{p}}(\me^{-x}x^{p+\al})
\end{equation}
We therefore suppose that
\[
F_{\ell m}(x) = \sum_{p}a_{\ell m p}f_{\ell p}(x)
\]
with
\[
f_{\ell p}(x) \equiv x^{\ell/2} L_{p}^{\ell+1/2}(x)
\]
Notice that $m$ does not appear in $f$; the constant coefficient $a_{\ell m p}$ is included to allow for $m$-dependence of $F_{\ell m}$.  As before we will often suppress the indices.

There are two arguments favoring this form for the $f_{\ell p}$:  \begin{enumerate}\item The $f_{\ell p}$ are orthogonal with respect to the natural weight function $x^{1/2}\me^{-x}$; that is,
\[
\int_{0}^{\infty}\dif x\, x^{1/2}\me^{-x}f_{\ell p}f_{\ell q} = \delta_{pq}\frac{\Gamma(\ell + 3/2 + n)}{n!}
\]
This follows from the Laguerre orthogonality condition
\begin{equation}\label{lagorthog}
\int_{0}^{\infty}dx\, e^{-x} x^{\alpha}L_{n}^{\alpha}L_{m}^{\alpha} = \delta_{mn}\frac{\Gamma(\al + 1 + n)}{n!} 
 \end{equation}

\item For small $\ell$ and $p$, the $f_{\ell p}$ have simple physical interpretations. Thus $f_{0 0}$ and $f_{0 1}$ correspond respectively to perturbation of the density and temperature, while $f_{10}$ and $f_{11}$ correspond respectively to particle flow and heat flow.  Thus:
\begin{eqnarray*}
L_{0}^{1/2}(x) &=&1 \,\,( = L_{0}^{3/2})\\
L_{1}^{1/2}(x) &=& 3/2 - x  \\
L_{1}^{3/2}  &=& 5/2 -x
\end{eqnarray*}

\end{enumerate}
Note the orthogonality condition also implies
\begin{equation}
\label{or1}
\int_{0}^{\infty}dx\, e^{-x} x^{\alpha}L_{p}^{\alpha} =  \delta_{p0}\Gamma(1+ \alpha) 
\end{equation}
since $L_{0}^{\al} = 1$.

We first compute the Rosenbluth potential $H_{lm}$ for Laguerre polynomials, using (\ref{defL}). The abbreviation 
\[
h_{\ell p} \equiv \mathcal{L}[f_{\ell p}]
\]
is convenient. To begin we assume $p> 0$; then (\ref{or1}) shows that the first term on the right-hand side of (\ref{defL}) does not contribute, so we have 
\[
h_{\ell p} = \frac{\me^{x}}{2\ell+1}\int_{x}^{\infty}\dif x'\left[\left(\frac{x}{x'}\right)^{\ell/2} -\left(\frac{x'}{x}\right)^{(\ell+1)/2}\right]\me^{-x'} x'^{\ell/2} L_{p}^{\ell+1/2}(x')
\]
According to (\ref{deflp}),
\[
h_{\ell p} = \frac{1}{(2\ell+1)p!}\int_{x}^{\infty}\dif y \left(\frac{x^{\ell/2}}{y^{\ell+1/2}} - \frac{1}{x^{(\ell+1)/2}}\right)\frac{d^{p}}{dy^{p}}\left(\me^{-y}y^{p+\ell+1/2}\right)
\]
or, after integration by parts,
\[
h_{\ell p} = \frac{1}{p!}x^{\ell/2}\int_{x}^{\infty}\dif y\, y^{-\ell -3/2}\frac{d^{p-1}}{dy^{p-1}}\left(\me^{-y}y^{p+\ell+1/2}\right)
\]
Note here that there is no boundary term, and that the $2\ell + 1$ factor has cancelled.  The integrand can be expressed as
\[
y^{-\ell -3/2}\frac{d^{p-1}}{dy^{p-1}}\left(\me^{-y}y^{p+\ell+1/2}\right) = (p-1)!\me^{-y}L_{p-1}^{\ell + 3/2}(y)
\]
so we have
\[
h_{\ell p} = \frac{1}{p}x^{\ell/2}\int_{x}^{\infty}dy \, \me^{-y}L_{p-1}^{\ell + 3/2}
\]
This integral is known (p. 809 of \cite{G&S}): 
\[
h_{\ell p} =  \frac{1}{p}x^{\ell/2}\me^{-x}[L_{p-1}^{\ell + 3/2} - L_{p-2}^{\ell + 3/2}]
\]
and then the identity  (p.1001 of \cite{G&S}
\[
L_{p-1}^{\ell + 3/2} - L_{p-2}^{\ell + 3/2} = L_{p-1}^{\ell + 1/2}
\]
implies
\begin{equation}
\label{punch1}
h_{\ell p} = \frac{\me^{-x}}{p} x^{\ell/2}L_{p-1}^{\ell+1/2}
\end{equation}

We conclude that
\begin{equation}
\label{hfin}
\mathcal{L}[f_{\ell p}] =  \frac{\me^{-x}}{p}f_{\ell (p-1)}, \,\, \mbox{for}\, p \geq 1
\end{equation}

In the case $p=0$,
\[
f_{\ell 0} = x^{\ell/2}
\]
and one finds from (\ref{defL}) that
\begin{equation}
\label{hl0}
\mathcal{L}[f_{\ell 0}] = \frac{1}{2}x^{-(\ell+1)/2}\gamma(\ell+1/2,x)
\end{equation}
Here we recall (\ref{defgam}).

The Laguerre-projection of $G_{\ell m}$ is more complicated.  We begin with an identity analogous to (\ref{1overu}):
\begin{equation}
\label{uyy}
u = -\sum_{\ell, m}\frac{ 4\pi x_{>}^{1/2}}{4\ell^{2} - 1}\left(\frac{x_{<}}{x_{>}}\right)^{\ell/2}\left(1 - b_{\ell}\frac{x_{<}}{x_{>}}\right)Y^{*}_{\ell m}(\theta',\gamma')Y_{\ell m}(\theta,\gamma)
\end{equation}
with
\[
b_{\ell} \equiv \frac{\ell-\frac{1}{2}}{\ell+\frac{3}{2}}
\]
This gives
\[
G_{\ell m}=\mathcal{K}[F_{\ell m}]
\]
where $\mathcal{K}$ is the operator
\[
\mathcal{K}[f]\equiv \frac{-1}{4\ell^{2} - 1}\int_{0}^{\infty}\dif x'\, \me^{-x'} x'^{1/2}x_{>}^{(1-\ell)/2} x_{<}^{\ell/2}\left(1 - b_{\ell}\frac{x_{<}}{x_{>}}\right)f
\]
Thus
\begin{equation}
\label{kflp}
\mathcal{K}[f_{\ell p}] = \frac{-1}{4\ell^{2} - 1}\int_{0}^{\infty}\dif x'\, \me^{-x'} x'^{(\ell + 1)/2}x_{>}^{(1-\ell)/2} x_{<}^{\ell/2}\left(1 - b_{\ell}\frac{x_{<}}{x_{>}}\right)L_{p}^{\ell + 1/2}
\end{equation}
At this point we note from (\ref{psi}) that $G$ appears in the collision operator only through the expression
\[
2x(2x G_{xx} + G_{x}) = 4x^{3/2} \frac{d}{dx}\left(\sqrt{x}\frac{dG}{dx}\right)
\]
Therefore we need to evaluate
\[
4x^{3/2} \frac{d}{dx}\left(\sqrt{x}\frac{d}{dx}\right)\mathcal{K}[f_{\ell p}] = -\frac{4x^{3/2}}{4\ell^{2} - 1}J
\]
where
\[
J \equiv \frac{d}{dx}\left(\sqrt{x}\frac{dI}{dx}\right)
\]
and
\[
\begin{split}
I &= \int_{0}^{x}\dif y \,\me^{-y} L y^{\ell + 1/2}\left( x^{(1-\ell)/2} - b_{\ell}yx^{-(1+\ell)/2}\right)\\ &+ \int_{x}^{\infty}\dif y \,\me^{-y}L \left(x^{\ell/2}y - b_{\ell}x^{1 + \ell/2}\right)
\end{split}
\]
Here and below we denote the Laguerre polynomial $L_{p}^{\ell + 1/2}$ simply by $L$.  We have
\[
\begin{split}
I_{x} &= \int_{0}^{x}\dif y \,\me^{-y} L y^{\ell + 1/2} x^{-(1+\ell)/2}\left[\frac{1-\ell}{2} + b_{\ell}\left(\frac{1+\ell}{2}\right)\frac{y}{x}\right]\\ &+ \int_{x}^{\infty}\dif y \,\me^{-y}L x^{\ell/2}\left[\frac{\ell y}{2x} - b_{\ell}\left(1 + \frac{\ell}{2}\right)\right]
\end{split}
\]
Note that the two endpoint contributions cancel.  From this point it is straightforward to compute
\begin{equation}
\label{jfin}
\begin{split}
J &= -x^{-(\ell+2)/2}\int_{0}^{x}\dif y \,\me^{-y} L\, y^{\ell + 1/2} \beta_{-}(x,y)\\ &- x^{(\ell - 3)/2}\int_{x}^{\infty}\dif y \,\me^{-y}L\, y\beta_{+}(x,y)
\end{split}
\end{equation}
Here we have introduced the abbreviations
\begin{eqnarray}
\bt_{-} &=&C_{0} + C_{1}\frac{y}{x} \\
\bt_{+} &=&C_{0} + C_{1}\frac{x}{y}
\end{eqnarray}
where
\begin{equation}
\label{defc12}
C_{0} = \frac{\ell}{2}\left(\frac{1-\ell}{2}\right), \,\, C_{1} = b_{\ell}\left(\frac{1+\ell}{2}\right)\left(1 + \frac{\ell}{2}\right)
\end{equation}

It is convenient to express (\ref{jfin}) as
\[
J = J_{0} +J_{+} +J_{-}
\]
where
\begin{eqnarray}
J_{0} &=& -x^{-(\ell + 2)/2}\int_{0}^{\infty}\dif y \, \me^{-y}L\,y^{\ell + 1/2}\bt_{-}(x,y) \\
J_{+} &=&  -x^{(\ell - 3)/2}\int_{x}^{\infty}\dif y \,\me^{-y}L\, y\beta_{+}(x,y) \\
J_{-} &=& x^{-(\ell +2)/2} \int_{x}^{\infty}\dif y \, \me^{-y}L y^{\ell + 1/2}\bt_{-}(x,y)
\end{eqnarray}

%
We consider first $J_{0}$, using (\ref{deflp}) to write
\[
J_{0} = \frac{x^{-(\ell + 2)/2}}{p!} \int_{0}^{\infty} \dif y \, (C_{0} + C_{1}y/x)\left(\frac{d}{dy}\right)^{p}(\me^{-y}y^{p + \ell + 1/2})
\]
The integral can be seen to vanish if $p>1$.  When $p= 0$, the Laguerre function is unity and we quickly find
\[
J_{0}(p = 0) = -x^{-(\ell + 2)/2}[C_{0}\Gamma(\ell + 3/2) + C_{1}x^{-1}\Gamma(\ell + 5/2)]
\]
When $p=1$,
\[
J_{0}(p = 1) =  -x^{-(\ell + 2)/2} x^{-1}C_{1}\int_{0}^{\infty}\dif y \, y \frac{d}{dy}(\me^{-y}y^{\ell + 3/2})
\]
integration by parts yields
\[
J_{0}(p = 1) =  x^{-(\ell + 2)/2} x^{-1}C_{1}\Gamma(\ell + 5/2)
\]
We conclude that
\begin{equation}
\label{j0fin}
J_{0} = -x^{-(\ell + 2)/2}[C_{0}\Gamma(\ell + 3/2) \delta_{p0} + x^{-1}C_{1}\Gamma(\ell + 5/2)(\delta_{p1} - \delta_{p0})]
\end{equation}

Consider next
\[
J_{+} =   -x^{(\ell - 3)/2}\int_{x}^{\infty}\dif y \,\me^{-y}L_{p}^{\ell + 1/2}\,(C_{0}y + C_{1}x)
\]
Regarding the $C_{1}$-term, we use the previously cited identity \cite{G&S}:
\begin{equation}
\label{ab1}
\int_{x}^{\infty}\dif y \, \me^{-y}L_{n}^{\al}(y) = \me^{-x}L_{n}^{\al - 1}(x)
\end{equation}
For the $C_{0}$ term, we use another identity \cite{G&S}:
\begin{equation}
\label{ab2}
yL_{p}^{\ell + 1/2} = (p + \ell + 1/2)L_{p}^{\ell - 1/2} - (p+1)L_{p+1}^{\ell - 1/2}
\end{equation}
Thus we find
\begin{equation}
\begin{split}
J_{+} = -x^{\ell - 3/2}\me^{-x}&\{C_{0} [(p + \ell + 1/2)L_{p}^{\ell - 3/2} - (p+1)L_{p+1}^{\ell - 3/2}] \\
+ C_{1}L_{p}^{\ell - 1/2}&\}
\end{split}
\end{equation}

Finally we compute $J_{-}$.  With the abbreviation $f = \me^{-y} y^{p+\ell +1/2}$, (\ref{deflp}) gives
\[
J_{-} = \frac{x^{-(\ell + 2)/2}}{p!} \int_{x}^{\infty}\dif y \, \left(C_{0} + C_{1}\frac{y}{x}\right)\left(\frac{d}{dy}\right)^{p}f(y) 
\]
It can be seen that the cases $p = 0$ and $p=1$ are exceptional.  We first assume $p\geq 2$ and infer, from (\ref{deflp}), 
\[
\left(\frac{d}{dy}\right)^{p-1}f(y) = (p-1)!L_{p-1}^{\al}\me^{-x}x^{\al}
\]
It is then straightforward to derive
\begin{equation}
\label{j-2}
J_{-} = -\me^{-x}\frac{x^{3(\ell + 1)/2}}{p}\left[C_{2} L_{p-1}^{\ell + 1/2} - C_{1}\frac{L_{p-2}^{\ell + 1/2}}{(p-1)} \right]
\end{equation}
Here we have introduced the constant
\begin{equation}
\label{defc}
C_{2} \equiv C_{0} + C_{1} = \frac{\ell + 1}{2}\left(\frac{2\ell - 1}{2\ell + 3}\right)
\end{equation}

The calculation for $p<2$ is similar, but simpler. The results combine neatly with the known form of $J_{0}$; one finds
\[
J_{0}(p= 0) + J_{-}(p=0) = -x^{-(\ell + 2)/2}\left[C_{0}\gamma(\ell + 3/2,x) + C_{1}x^{-1}\gamma(\ell + 5/2,x)\right]
\]
and 
\[
J_{0}(p=1)+ J_{-}(p = 1) = -x^{-(\ell + 2)/2}\left[C_{2}\me^{-x}x^{\ell + 3/2} - C_{1}x^{-1}\gamma(\ell + 5/2,x)\right]
\]
\newpage

\bibliographystyle{plain}

\begin{thebibliography}{1}

\bibitem{abram}
Milton Abramowitz and Irene~A. Stegun.
\newblock {\em Handbook of Mathematical Functions}.
\newblock National Bureau of Standards, 1964.

\bibitem{braginskii}
S.~I. Braginskii.
\newblock Transport processes in a plasma.
\newblock In M.~A. Leontovich, editor, {\em Reviews of Plasma Physics},
  volume~1. Consultants Bureau, New York, 1965.

\bibitem{G&S}
I.~S. Gradshteyn and I.~M. Ryzhik.
\newblock {\em Tables of Integrals, Series and Products}.
\newblock Academic Press, New York, 1965.

\bibitem{meiss92}
R.~D. Hazeltine and J.~D. Meiss.
\newblock {\em Plasma Confinement}.
\newblock Frontiers in Physics. Addison-Wesley, Redwood City, 1992.

\bibitem{framework}
R.~D. Hazeltine and F.~L. Waelbroeck.
\newblock {\em The Framework of Plasma Physics}.
\newblock Perseus Books, Reading, Massachusetts, 1998.

\bibitem{RMJ}
M.~N. Rosenbluth, William~M. MacDonald, and David~L. Judd.
\newblock Fokker-{P}lanck equation for an inverse-square force.
\newblock {\em Physical Review}, 107:1--6, 1957.

\end{thebibliography}
\end{document}